\documentclass[12pt,a4paper]{article}
\pdfoutput=1
\usepackage{jheppub}
\usepackage{feynmp-auto}
\usepackage[utf8]{inputenc}
\usepackage[fleqn]{mathtools}
\usepackage[fleqn]{amsmath}
\usepackage{amssymb,mathrsfs}
\usepackage{setspace}
\usepackage{graphicx,floatflt,rotate, cancel}
\usepackage{bm}
\usepackage{color}
\usepackage{xcolor}
\usepackage{graphicx}
\usepackage{multirow}
\usepackage{array}
\usepackage{float}
\usepackage[toc,page]{appendix}
\usepackage{diagbox}
\usepackage[mathscr]{euscript}
\usepackage{cleveref}
\usepackage{soul}
\usepackage{pifont}
\usepackage[normalem]{ulem}
\usepackage[numbers]{natbib}
\usepackage{notoccite}
\usepackage{comment}
\makeatletter
\gdef\@fpheader{}
\makeatother

\definecolor{orange}{rgb}{1,0.5,0}

\title{Vector Dark Matter in a $U(1)_X$ extended 2HDM}
\author[a]{Nandini Das,}
\author[b,c]{Juhi Dutta,}
\author[a]{Dilip Kumar Ghosh,}
\author[d]{Santosh Kumar Rai}
\affiliation[a]{School of Physical Sciences, Indian Association for the Cultivation of Science, 2A $\&$ 2B, Raja S.C. Mullick Road, Kolkata 700032, India}
\affiliation[b]{Homer L. Dodge Department of Physics and Astronomy,
University of Oklahoma, Norman, OK 73019, USA}
\affiliation[c]{The Institute of Mathematical Sciences, 
4th Cross St, CIT Campus, Tharamani, Chennai, Tamil Nadu, India, 600113}
\affiliation[d]{Regional Centre for Accelerator-based Particle Physics, Harish-Chandra Research Institute, A CI of Homi Bhabha National Institute, Chhatnag Road, Jhunsi, Prayagraj – 211019, India}

\emailAdd{nandinidas.rs@gmail.com}
\emailAdd{juhidutta@imsc.res.in}
\emailAdd{tpdkg@iacs.res.in}
\emailAdd{skrai@hri.res.in}

\abstract{
We investigate the possibility of having a vector boson dark matter in a $U(1)_X$ extended two-Higgs-doublet model (2HDM) setup. The gauge boson gains mass when a SM singlet complex scalar, which is charged under the dark $U(1)_X$ symmetry, acquires a vacuum expectation value (\textit{vev}). This scalar acts as the connection between the SM sector and DM via the Higgs portal. An additional exact charge conjugation symmetry inhibits the mixing of this gauge boson with the photon, thereby confirming the stability of DM. On the other hand, 2HDM with Type I $Z_2$ restriction can offer a non-standard Higgs in the lighter mass range. This freedom allows us to accommodate dark matter mass in the (40-60) GeV regime where the direct detection constraints are strongest. We study the dark matter phenomenology of such a model while taking care of all possible theoretical and experimental constraints.}
\preprint{HRI-RECAPP-2025-05}
\begin{document}

\maketitle

\setcounter{footnote}{0}  
\renewcommand{\thefootnote}{\arabic{footnote}}  

\section{Introduction} \label{sec:intro}
The Standard Model (SM) is the most widely accepted description of the physics of electroweak and strong interactions. The recent discovery of the Higgs boson only strengthens its construction. Although SM is extremely successful in explaining the particle interaction of the visible sector, it remains silent about the ``invisible" component of the universe. 
From Fritz Zwicky’s observations of galaxy clusters \cite{Zwicky:1933gu}, to Vera Rubin’s studies of galactic rotation curves \cite{Rubin:1970zza}, and the Bullet Cluster data from the Chandra X-ray Observatory~\cite{Barrena:2002dp, Clowe:2006eq}, multiple observations have confirmed the presence of a mysterious ``dark" component of the universe, non-luminous and non-baryonic in nature.
Supported by the aforementioned various hints of evidence along with gravitational lensing and anisotropies of Cosmic Microwave Background Radiation (CMBR)~\cite{Planck:2018vyg}, dark matter (DM), which supplies $26\%$ of the energy budget of the universe, is a mystery yet to be unveiled. One of the known facts about DM is its current number density ($\Omega_{DM} h^2$), which has been measured by {\tt WMAP} \cite{WMAP:2006bqn} and {\tt PLANCK} \cite{Planck:2018vyg} very precisely and is reported to be $\Omega_{DM} h^2=0.12 \pm 0.001$ at $68\%$ confidence level. In other words, DM in terms of interaction, mass, or nature is an unsolved puzzle. Although evidence only reveals its gravitational nature, exploring its particle origin is still an interesting and complementary possibility. From the analogy of the visible universe, if DM is considered an elementary particle, it should be connected to the SM by some interaction, most likely a weak interaction. Interestingly, weakly interacting particles of an electroweak mass scale can be in thermal equilibrium with the SM particle bath in the early universe and can freeze out from the bath with a residual number density that matches the observed DM density of the universe. This is often referred to as the WIMP miracle \cite{Kolb:1990vq} in the literature. Due to its detection prospects, WIMP is the most economical dark matter candidate to date. Multiple 
direct detection (DD) experiments like {\tt LUX-ZEPLIN} (LZ) \cite{LZ:2019sgr,LZ:2022lsv,LZ:2024zvo},(the most stringent limit till now), {\tt PANDAX} \cite{PandaX-II:2016vec,PandaX-II:2017hlx}, and {\tt XENONnT} \cite{XENON:2018voc,XENON:2023cxc} have explored and excluded wide ranges of parameter spaces. Futuristic experiments such as the {\tt DARWIN} \cite{DARWIN:2016hyl} intend to probe further parameter spaces in the coming years. Current direct search experiments have stringently constrained WIMP DM scenarios, particularly scalar and fermionic DM, which have been extensively looked into in various beyond Standard Model (BSM) theories. However, vector DM candidates have been relatively less explored.
Vector DM has been studied mainly in the context of SM extensions via the Higgs portal with extra scalars~\cite{Hambye:2008bq, Hambye:2009fg, Arina:2009uq, Lebedev:2011iq,Abe:2012hb, Farzan:2012hh,Choi:2013eua,Bernal:2015ova,Bhattacharya:2021rwh, YaserAyazi:2024hpj,Yamashita:2024krp,Khan:2025keb}, including $SU(2)_N$ extensions~\cite{Barman:2017yzr,Biswas:2018sib,Tran:2023lzv,Covi:2025erx}, vector-like fermion portals~\cite{Belyaev:2022shr}, and Type II seesaw~\cite{Das:2024tfe}.  

On the other hand, the two-Higgs doublet model (2HDM) \cite{Branco:2011iw} is a very well-motivated extension of the SM. 
Beyond its minimal scalar sector with two $SU(2)_L$ Higgs doublets, which can provide sources of spontaneous CP violation \cite{Gunion:2005ja}, explain the fermion mass hierarchy, etc., it can also provide an alternate perspective towards dark matter modeling. However, except for the inert Higgs doublet model \cite{LopezHonorez:2006gr}, where the non-standard Higgs doublet does not acquire a VEV and can play the role of dark matter, only the 2HDM can not have a natural dark matter candidate.
 
In this context, extended 2HDMs have been widely studied in the literature addressing the dark matter issue with fermion \cite{Arcadi:2018pfo, PhysRevD.89.033007, Arcadi_2018, CAMARGO2019319,Das:2024xle}, pseudo-Nambu Goldstone boson \cite{PhysRevD.100.075011, Biekotter:2022bxp,Jiang:2023xdf, Biekotter:2021ovi} and scalar particle \cite{Aoki:2009pf, Okada:2014usa, Okada:2013bna, Bonilla:2014xba,Goudelis:2013uca, PhysRevD.80.055012, Drozd:2014yla, Arhrib:2018qmw, Dey:2019lyr,Dutta:2022qeq,Herms:2022nhd,Dutta:2023cig,Dutta:2025nmy, Abdallah:2024npf}. Several searches \cite{Cai:2013zga,Chang:2017gla,Bertuzzo:2024bwy} have been conducted following the effective field theory approach, where the phenomenology of fermion, scalar, and vector boson as candidates for DM has been investigated. However, vector boson DM in a UV-complete 2HDM model framework has not been explored, and we focus on that scenario here.

In this work, we study the phenomenology of a vector dark matter candidate in the context of $U(1)_X$ extended Type I two Higgs doublet model (2HDM). The choice of Type I is motivated by the relatively relaxed constraints of the non-standard Higgs. Along with the two $SU(2)_L$ Higgs doublets, an additional complex scalar is present in the particle spectrum, which is charged under an extra local $U(1)_X$ symmetry. This $U(1)_X$ is spontaneously broken by the VEV of the complex scalar, while the DM gauge boson acquires mass by eating up the CP-odd part of the complex scalar. The dark matter is stabilized by the imposition of an exact charge conjugation symmetry. Therefore, after electroweak symmetry breaking, the particle content consists of three CP-even Higgs fields $h_1,h_2,h_3$, one CP-odd pseudoscalar $A$, a pair of charged Higgses $H^{\pm}$, and a gauge boson DM candidate $Z^{\prime}$. Taking care of all possible theoretical and experimental constraints, we scrutinize our DM model parameter space under the existing and future direct detection experimental facilities.   We provide the corresponding DM parameter space that can be probed soon by current and future experiments, such as DARWIN. 

It is worth mentioning here that the presence of an extra Higgs doublet along with a complex singlet scalar in the particle spectrum could be motivated in multiple ways, as noted below:
\begin{enumerate}
\item The choice of a light scalar in a model, which is not yet excluded by experimental observations, is rather limited. The Type I 2HDM still allows such a light non-standard scalar, which, being in the $[80-120]$ GeV mass range, can contribute to the annihilation of a light DM of $[40-60]$ GeV. 
In the scenario considered here, where the $U(1)_X$ scalar is assumed to be heavy, the light non-standard Higgs doublet plays a crucial role in supporting a viable light-DM parameter space.
 
 \item Additionally, the three light scalars ($H^{\pm}, h_1, A$) also help the dark matter to achieve thermal equilibrium in the early Universe.

    \item  Introducing an additional non-standard Higgs doublet into the scalar sector significantly expands the viable dark matter parameter space compared to the minimal vector-like DM scenario. A similar comparison of the parameter space has been studied in Ref.~\cite{Das:2024tfe} for the case of an SU(2)$_L$ triplet scalar. We anticipate comparable 
DM phenomenology in the present doublet extension. Nevertheless, the key advantage exploited in this work is the relative lightness 
of extra scalars $(h_1, A, H^\pm) $, which opens up a light-DM regime that remained inaccessible in the triplet-based scenario of Ref.\cite{Das:2024tfe}.

\end{enumerate}

This paper is organized as follows: We describe the model setup briefly in Section \ref{sec:model}. The theoretical and experimental constraints are summarized in Section \ref{sec:constraints} and their impact on the parameter space is shown in Section \ref{sec:imconstraints}. The DM phenomenology of the model, considering the freeze-out mechanism of production and its detectability, 
is shown in Section \ref{sec:DM}. The freeze-in prospect of the model is discussed in Section \ref{sec:Freezein}. Finally, we conclude in Section \ref{sec:summary}.

\section{Model Setup} \label{sec:model}
The SM gauge symmetry is extended with an additional $U(1)_X$ symmetry. In addition to the conventional content of the two-Higgs doublet model (which extends the SM by an extra SU(2)$_L$ Higgs doublet), we have an extra singlet scalar which is charged under the new $U(1)_X$ symmetry. Note that the new gauge symmetry represents a dark sector, and the gauge boson, which arises from gauging the symmetry, acts as the dark matter candidate in the model. The model Lagrangian can be written as
\begin{equation*}
    \mathcal{L}=\mathcal{L^{{\rm 2HDM}+ \rm Singlet}_{\rm Scalar}}+\mathcal{L^{Z^\prime}_{\rm Dark}}+\mathcal{L^{\rm 2HDM}_{\rm Yukawa}}+\mathcal{L^{\rm 2HDM}_{\rm Gauge}}
\end{equation*} 
where the first term corresponds to the scalar sector of the model, the second term corresponds to the dark sector, the third term corresponds to the Yukawa sector, and the last term to the gauge sector of the SM. We describe each term in detail below. 

\subsection{Scalar Sector} 
The most general Lagrangian including two Higgs doublets and a $U(1)_X$ charged singlet scalar can be written as
\begin{equation*}
    \mathcal{L^{{\rm 2HDM}+ \rm Singlet}_{\rm Scalar}}= \mathcal{L_{\rm Kin}}-V(\Phi_1, \Phi_2, S)
\end{equation*}
where $\Phi_1,\Phi_2$ and $S$ are the two Higgs doublets and the complex singlet scalar, respectively.

The kinetic part of the Lagrangian comprises
\begin{equation*}
    \mathcal{L_{\rm Kin}}=(D_{\mu} \Phi_1)^\dagger (D^{\mu} \Phi_1) + (D_{\mu} \Phi_2)^\dagger (D^{\mu} \Phi_2) + (D_{\mu} S)^* (D^{\mu} S) 
\end{equation*}
where the covariant derivatives are as follows 
\begin{eqnarray*}
  D_{\mu} \Phi_1 &=& \partial_\mu \Phi_1 + \frac{i g}{2} W_{\mu}^a \sigma^a \Phi_1+ \frac{i g^\prime}{2} B_{\mu} \Phi_1 \\
   D_{\mu} \Phi_2 &=& \partial_\mu \Phi_2 + \frac{i g}{2} W_{\mu}^a \sigma^a \Phi_2+ \frac{i g^\prime}{2} B_{\mu} \Phi_2 \\
   D_{\mu} S &=& \partial_\mu S + i g_x x_s Z^\prime_\mu S.
    \end{eqnarray*}
    Here, $g_x$ and $x_s$ are the gauge coupling and charge of the scalar $S$ under the dark $U(1)_X$, respectively. For simplicity, we choose $x_s=1$ for the remainder of the study.
    
The scalar potential is given by
\begin{eqnarray*}
    V(\Phi_1, \Phi_2, S)&=& m^2_{11} (\Phi_1^\dagger \Phi_1)+m^2_{22} (\Phi_2^\dagger \Phi_2)+ m^2_{33} (S^* S) -[m^2_{12} (\Phi_1^\dagger \Phi_2)+{\rm h.c.}]+  \lambda_1 (\Phi_1^\dagger \Phi_1)^2  \\
     &+& \lambda_2 (\Phi_2^\dagger \Phi_2)^2 +\lambda_S (S^* S)^2+ \lambda_3 (\Phi_1^\dagger \Phi_1) (\Phi_2^\dagger \Phi_2)+\lambda_4 (\Phi_1^\dagger \Phi_2) (\Phi_2^\dagger \Phi_1)\\
     &+& \lambda_{S1} (\Phi_1^\dagger \Phi_1) (S^* S)+ \lambda_{S2} (\Phi_2^\dagger \Phi_2) (S^* S) +\Big\{\frac{1}{2} \lambda_5 (\Phi_1^\dagger \Phi_2)^2 \\
     &+& [\lambda_6 (\Phi_1^\dagger \Phi_1)+\lambda_7 (\Phi_2^\dagger \Phi_2)+ \lambda_8 (S^* S)]
 \times (\Phi_1^\dagger \Phi_2) +\rm {h.c.}\Big\}
 \label{eq:scpot}
    \end{eqnarray*}
    where h.c. stands for the hermitian conjugate of the corresponding term. Following the $Z_2$ symmetry of Type I 2HDM, under which one of the Higgs doublets (in our case $\Phi_2$) is odd, $\lambda_6$, $\lambda_7$, and $\lambda_8$ must vanish to avoid explicit $Z_2$ breaking terms. However,  $m^2_{12}$ can have non-zero values and appear as a soft $Z_2\!\!\!\!\!/$ term. After electroweak symmetry breaking, the two scalar doublets $\Phi_1$ and $\Phi_2$ and the singlet scalar $S$ can be expanded around the \textit{vev's} as 
    \begin{eqnarray*}
      \Phi_1 &=& \begin{pmatrix}
      \phi^{\dagger}_1 \\
      \frac{1}{\sqrt{2}}(\rho_1 +v_1 + i \eta_1)
          \end{pmatrix},
         ~~  \Phi_2 = \begin{pmatrix}
      \phi^{\dagger}_2 \\
      \frac{1}{\sqrt{2}}(\rho_2 +v_2 + i \eta_2)
          \end{pmatrix},\\
          &&~~~~~~~~~~~~~~~S = \frac{1}{\sqrt{2}} (\rho_S + v_S + i \eta_S)
    \end{eqnarray*}
    where $v_1,v_2$ and $v_S$ are the $vev's$ of the two Higgs doublets and the singlet scalar, respectively.
    Minimizing the scalar potential around the \textit{vev}, we obtain the following relations,
\begin{eqnarray*}
  m^2_{11} &=& m^2_{12} \frac{v_2}{v_1}- \lambda_1 v^2_1 - \frac{1}{2}(\lambda_3+\lambda_4+\lambda_5) v^2_2- \frac{\lambda_{S1}}{2} v^2_S \\
 m^2_{22} &=& m^2_{12} \frac{v_1}{v_2}- \lambda_2 v^2_2 - \frac{1}{2}(\lambda_3+\lambda_4+\lambda_5) v^2_1- \frac{\lambda_{S2}}{2} v^2_S \\
-m^2_{33}&=&\frac{\lambda_{S1}}{2} v^2_1 +\frac{\lambda_{S2}}{2} v^2_2 +\lambda_{S} v^2_S \,\, .
\end{eqnarray*}

   The mass matrix of the charged scalars in the gauge eigen basis is given by
   \begin{equation}
    \mathcal{M}^2_{\pm}= \begin{pmatrix}
        m^2_{12} \frac{v_2}{v_1}-\frac{v^2_2}{2}(\lambda_4+\lambda_5)    &     -m^2_{12}+\frac{1}{2} v_1 v_2 (\lambda_4+ \lambda_5) \\
    -m^2_{12}+\frac{1}{2} v_1 v_2 (\lambda_4+ \lambda_5) &   m^2_{12} \frac{v_1}{v_2}-\frac{v^2_1}{2}(\lambda_4+\lambda_5)   
    \end{pmatrix} \,\, .
   \end{equation}

The field rotation of the charged scalars is given by 
   \begin{eqnarray*}
       \begin{pmatrix}
           \phi_1^{\pm}\\
           \phi_2^{\pm}
       \end{pmatrix}
       &=& \begin{pmatrix}
           {\rm cos\beta} & {\rm sin\beta} \\
           {\rm sin\beta} & -{\rm cos\beta} \\
           \end{pmatrix}
           \begin{pmatrix}
               G^{\pm} \\
               H^{\pm}
           \end{pmatrix}
   \end{eqnarray*}
   where 
   \begin{equation}
     {\rm tan \beta}=\frac{v_2}{v_1} \,\, .
   \end{equation}
   One of the charged mass eigenstates becomes the Goldstone boson corresponding to the $W^{\pm}$ boson. The mass eigenvalue of the other charged scalar is given by
    \begin{equation}
        m^2_{H^{\pm}}=(\frac{m^2_{12}}{v_1 v_2}-\frac{1}{2}(\lambda_4+\lambda_5)) (v^2_1+ v^2_2) \,\, .
    \end{equation}
    The mass matrix of the CP-odd scalars in the gauge eigen basis is given by 
    \begin{equation}
       \mathcal{M}^2_O=\begin{pmatrix}
           m^2_{12}\frac{v_2}{v_1}-\lambda_5 v^2_2  & -m^2_{12}+v_1 v_2 \lambda_5 \\
           m^2_{12}+v_1 v_2 \lambda_5 &  m^2_{12}\frac{v_1}{v_2}-\lambda_5 v^2_1
           
       \end{pmatrix}  \,\, .
    \end{equation}
    The relation between the gauge basis and mass basis of the CP odd scalars can be written as 
    \begin{eqnarray*}
       \begin{pmatrix}
           \eta_1\\
           \eta_2
       \end{pmatrix}
       &=& \begin{pmatrix}
           {\rm cos\beta} & {\rm sin\beta} \\
           {\rm sin\beta} & -{\rm cos\beta} \\
           \end{pmatrix}
           \begin{pmatrix}
               G_0 \\
               A
           \end{pmatrix} \,\, .
   \end{eqnarray*}
   One of the pseudo-scalar acts as a Goldstone boson of the $Z$ boson, whereas the mass eigenvalue of the other pseudo-scalar is given by 
   \begin{equation}
       m^2_{A}=\frac{(m^2_{12}-\lambda_5 v_1 v_2)}{v_1 v_2} (v^2_1+ v^2_2) \,\, .
   \end{equation}
   The mass matrix of the CP-even scalars is given by
   \begin{equation}
       \mathcal{M}^2_{E}=\begin{pmatrix}
           m^2_{12} \frac{v_2}{v_1}+2 \lambda_1 v^2_1 & -m^2_{12}+\lambda_{345} v_1 v_2  & \lambda_{S1} v_1 v_S   \\
           -m^2_{12}+\lambda_{345} v_1 v_2 & m^2_{12} \frac{v_1}{v_2}+2 \lambda_2 v^2_2 & \lambda_{S2} v_2 v_S \\
           \lambda_{S1} v_1 v_S  & \lambda_{S2} v_2 v_S  & 2 \lambda_{s} v^2_S
           \end{pmatrix}
   \end{equation}
   where 
\begin{equation*}
    \lambda_{345}=\lambda_3+\lambda_4+\lambda_5.
\end{equation*}
    The CP even gauge eigenstates can be written as 
    \begin{eqnarray}
        \begin{pmatrix}
            \rho_1\\
            \rho_2 \\
            \rho_S
        \end{pmatrix}&=& 
        \begin{pmatrix}
                c_{\alpha_1} c_{\alpha_2} &  -s_{\alpha_1} c_{\alpha_3} - c_{\alpha_1} s_{\alpha_2} s_{\alpha_3}  & s_{\alpha_1} s_{\alpha_3} - c_{\alpha_1} c_{\alpha_3} s_{\alpha_2} \\
                c_{\alpha_2} s_{\alpha_1} 
                &  c_{\alpha_1} c_{\alpha_3}- s_{\alpha_1} s_{\alpha_2} s_{\alpha_3}  & -(c_{\alpha_1} s_{\alpha_3}+ c_{\alpha_3} s_{\alpha_1} s_{\alpha_2})\\
                s_{\alpha_2} & c_{\alpha_2} s_{\alpha_3} & c_{\alpha_2} c_{\alpha_3}
        \end{pmatrix} 
        \begin{pmatrix}
            h_1 \\
            h_2 \\
            h_3
        \end{pmatrix}
        \label{eq:CPH}
    \end{eqnarray}
    where $h_1, h_2$ and $h_3$ are the mass eigenstates and $\alpha_1,\alpha_2,\alpha_3$ are the mixing angles in the CP-even Higgs sector.  For our future references, we denote the $3\times 3 $ rotation matrix shown in eq.\ref{eq:CPH} as ${\cal R}$, whose elements will play a 
    crucial role in various Higgs couplings. The quartic couplings can be written in terms of the physical parameters (mass, mixing angles, and vev). The expressions are given in the 
    Appendix \ref{qcoup}. After trading off the three unphysical mass parameters by using the minimization conditions, we are left with the following set of free parameters
\begin{equation}
    \{ m_{h_1}, m_{h_2}, m_{h_3}, m_A, m_{H^{\pm}}, \sin\alpha_{1}, \sin\alpha_{2}, \sin\alpha_{3}, \tan\beta, m^2_{12}, v_S\}
\end{equation}
where $m_{h_1},m_{h_2},m_{h_3}$ are the mass eigenvalues of the CP-even neutral scalars, $m_A$ and $m_{H^{\pm}}$ are the mass eigenvalues for the CP-odd neutral scalar and the singly charged scalar, respectively. 
  
\subsection{Yukawa Sector}  
   For a general 2HDM, where no $Z_2$ symmetry is imposed on the Lagrangian, we can write the Yukawa terms of the Lagrangian as
    \begin{eqnarray*}
     -\mathcal{L}_{{\rm Yukawa}} &=& \overline{Q}_{Li} ({Y^d_1}_{ij} \Phi_1+{Y^d_2}_{ij} \Phi_2) d_{Rj} + \overline{Q}_{Li} ({Y^u_1}_{ij} \tilde{\Phi}_1+{Y^u_2}_{ij} \tilde{\Phi}_2) u_{Rj} \\
     &+&
        \overline{L}_{Li} ({Y^{\ell}_1}_{ij} \Phi_1+ {Y^{\ell}_2}_{ij} \Phi_2)e_{Rj} +h.c.   
    \end{eqnarray*}
where $Y^f_\alpha$ corresponds to the Yukawa couplings of $f\,(f=u,d,\ell)$ type fermions to the $\Phi_\alpha (\alpha=1,2)$ Higgs doublet and $i$ and $j$ stand for fermion generation indices. After imposition of the $Z_2$ symmetry of Type I 2HDM on our Yukawa Lagrangian, the fermions only couple to the second Higgs doublet ($\Phi_2$), which is dominated by the SM Higgs component on the mass basis. The corresponding Lagrangian for Type I 2HDM is
\begin{equation}
        -\mathcal{L}^{2HDM}_{{\rm Yukawa}} = \overline{Q}_{Li} {Y^d_2}_{ij} \Phi_2 d_{Rj} + \overline{Q}_{Li} {Y^u_2}_{ij} \tilde{\Phi}_2 u_{Rj} +
        \overline{L}_{Li}  {Y^{\ell}_2}_{ij} \Phi_2e_{Rj} + h.c.  
      \end{equation}
 In the mass basis of Higgs and fermions, the Lagrangian can be written as~\cite{Branco:2011iw}
\begin{eqnarray*}
   -\mathcal{L}_{\rm Yuk} &=&  - \sum_{f=u,d,\ell} \frac{m_f}{v} (\xi_{h_i}^f \bar{f}f h_i+- i \xi_A^f  \bar{f}\gamma_5f A
 )- \Big(\frac{\sqrt{2}V_{ud}}{v}\, \bar{u} \left( m_u \xi_A^u \text{P}_L+ m_d \xi_A^d \text{P}_R \right)d H^+\Big)\\
&&+\frac{\sqrt2m_\ell\xi_A^\ell}{v}\, \bar{\nu}_L^{}\ell_R^{}H^+
 +\text{H.c.}
\end{eqnarray*}
where $m_f$ stands for the fermion mass. The modified reduced coupling factors of the CP even Higgses to the fermions, $\xi_{h_i}^f$ are defined as~\cite{Heinemeyer:2021msz}: 
\begin{align}
    \xi_{h_i}^f&= \frac{\mathcal{R}_{i2}}{\sin \beta} 
\end{align}
where $i$ runs from 1 to 3, corresponding to three CP even Higgses. $\mathcal{R}_{i2}$ ($i=1,2,3$) are the elements of the rotation matrix in the CP-even Higgs sector, as shown in eq.~\ref{eq:CPH}. The reduced couplings of the CP odd scalar ($\xi^f_A$) and the charged scalar to the fermions are the same as the standard Type I 2HDM~\cite{Branco:2011iw}. The alignment limit for this extended version of 2HDM is $\{\beta - (\alpha_1 + \text{sgn}(\alpha_2)\alpha_3)\}\rightarrow \pi /2$ for the case where $h_2$ acts as the SM-like Higgs boson. In the limit $\sin\alpha_2 \to 0~ \& \sin\alpha_3 \to 0$, the CP even reduced couplings return to the standard Type I 2HDM couplings.

\subsection{Dark Sector}
\color{black} 
The $U(1)_X$ symmetric Lagrangian is invariant under $Z^{(A)}_2 \times Z^{(B)}_2$ symmetry~\cite{Farzan:2012hh}  which are defined as follows
\begin{equation*}
Z^{(A)}_2:~~~ Z^\prime_{\mu} \to -Z^\prime_{\mu}, S \to S^*,~~~
Z^{(B)}_2:~~~ Z^\prime_{\mu} \to -Z^\prime_{\mu}, S \to -S^*
\end{equation*}
 This is also referred to as dark charge conjugation symmetry~\cite{Ma:2017ucp}. 
Here we would like to clarify that the $Z_2$ symmetries are not put in by hand in this scenario. The $U(1)_X$ symmetric potential has the symmetry $S \to S^*$ or $S \to -S^*$. To make the kinetic term of $S$ invariant under the above symmetry operation, the transformation of $Z^\prime$ has to be  $Z^\prime_{\mu} \to -Z^\prime_{\mu}$. These two sets of symmetry transformations are defined as $Z^{(A)}_2$ and $Z^{(B)}_2$. If any of these symmetries is exactly respected, the kinetic mixing term of the $U(1)$ gauge fields has to be avoided. which consequently ensures the stability of  the vector boson $Z^\prime$.
 In the presence of the dark conjugation symmetry, the neutral gauge boson mass matrix is given by
\begin{equation}
M^2_{VZ} =  
\begin{pmatrix}
     g^2\frac{v^2}{4}  & -gg^{\prime}\frac{v^2}{4} & 0\\
   -gg^{\prime}\frac{v^2}{4} &   g^{\prime2}\frac{v^2}{4}  & 0\\
    0 & 0 &   g^2_xv^2_S \\
\end{pmatrix}.
\end{equation}  
Upon diagonalizing the matrix, the eigenvalues are
$0, \frac{(g^2+g^{\prime2})v^2}{2}, g^2_xv^2_S$ representing the massless photon, the massive $Z$ boson, and the massive $Z^{\prime}$ boson. The  $Z^{\prime}$ boson  constitutes the  vector DM candidate such that 
\begin{equation}
    m_{Z^{\prime}}=g_xv_S
\end{equation} 
where $m_{Z^{\prime}}$ is the mass of the $Z^{\prime}$ vector boson DM candidate. 
 In our analysis, we treat $m_{Z^\prime}$ and
$g_x$ as free parameters. For the remainder of the study, we define the DM mass as $m_{DM}$, where $m_{DM}=m_{Z^{\prime}}$.
Consequently, incorporating the dark sector, the model comprises twelve free parameters, which are listed as follows: 

\begin{eqnarray} 
m_{h_1},m_{h_2},m_{h_3},m_{H^{\pm}}, m_{A},m^2_{12}, \tan \beta, \sin \alpha_1,\sin \alpha_2,\sin \alpha_3,  m_{DM}, g_x
\end{eqnarray}
One of the parameters, $m_{h_2}$, which will be identified as the SM-like Higgs boson, is fixed at the experimentally observed value of 125 GeV. We assume degeneracy among $h_1, H^\pm$, and $A$ to further reduce the number of free parameters. This choice of degeneracy is motivated by constraints from electroweak precision measurements.

A vector boson as dark matter interacts with the SM sector via the Higgses. The vertex factors for the relevant trilinear and quartic couplings of dark matter are as follows 
\begin{eqnarray}
 \lambda_{h_i Z^{\prime} Z^{\prime}} = 2g^2_x v_S \mathcal{R}_{i3} \nonumber \\
    \lambda_{Z^{\prime} Z^{\prime}  h_i h_j} =2g^2_x\mathcal{R}_{i3}\mathcal{R}_{j3} 
    \label{eq:couplings}
\end{eqnarray}  
where $\mathcal{R}_{ij}$ ($i,j=1,2,3$) are the elements of the rotation matrix in the CP-even Higgs sector as shown in eq.~\ref{eq:CPH}.

\begin{figure}
    \centering
    \includegraphics[scale=0.75]{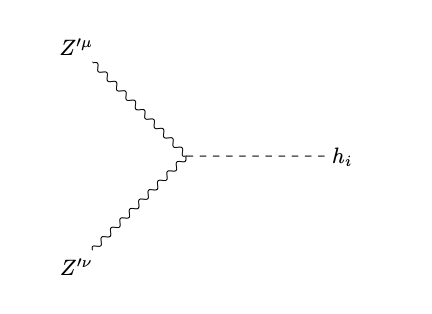}
    \includegraphics[scale=0.65]{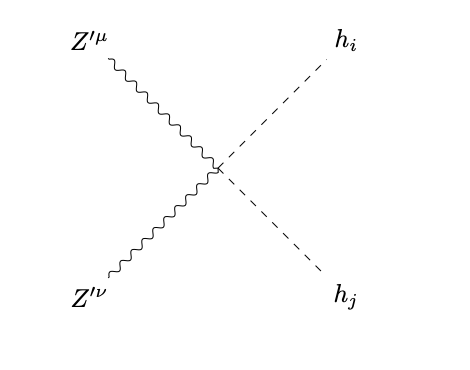}
    \caption{Relevant trilinear and quartic vertices of vector DM candidate $Z^{\prime}$.}
    \label{fig:coup}
\end{figure}
\color{black}
\section{Constraints} \label{sec:constraints}
\subsection{Theoretical constraints}
\subsection*{Constraints from boundedness-from-below conditions}

The conditions of boundedness-from-below (BFB) are crucial to ensure the positivity of the scalar potential, i.e., $V(\Phi_1,\Phi_2,S)\geq0$ for large values of the fields. Using copositivity and Sylvester criterion~\cite{Kannike:2012pe}, the necessary and sufficient BFB conditions are,
    \begin{eqnarray}
       \lambda_1 \geq0, ~~\lambda_2 \geq 0,~~\lambda_S \geq 0,
      \lambda_3+\lambda_4-|\lambda_5|+2 \sqrt{\lambda_1\lambda_2}\geq 0,\lambda_3+ 2 \sqrt{\lambda_1\lambda_2}\geq 0, \nonumber\\
      \lambda_{S1} +2\sqrt{\lambda_1\lambda_S}\geq0, \lambda_{S2} +2 \sqrt{\lambda_2\lambda_S}\geq0, \nonumber \\
       2\sqrt{2\lambda_1\lambda_2\lambda_S}  + \lambda_{S1}\sqrt{2\lambda_2}+ \lambda_{S2}\sqrt{2\lambda_1}+ 
       (\lambda_3+\lambda_4-|\lambda_5|)\sqrt{2\lambda_{S}}\nonumber\\+\sqrt{2(\lambda_{S1}+2\sqrt{\lambda_1\lambda_S})(\lambda_{S2}+2\sqrt{\lambda_2\lambda_S})((\lambda_3+\lambda_4-|\lambda_5|)+2\sqrt{\lambda_1\lambda_2})} \geq 0.\nonumber 
    \end{eqnarray}
In the limit where $\lambda_s \to 0$, $\lambda_{S1} \to 0$ and $\lambda_{S2} \to 0 $, the conditions reduce to the BFB conditions of standard 2HDM~\cite{Branco:2011iw}. 
\subsection*{Constraints from Unitarity}
The unitarity constraints for this model setup are as follows
\begin{eqnarray}
    \lambda_3+\lambda_4 < 8 \pi, ~~~~\lambda_3-\lambda_4 < 8 \pi, ~~~\lambda_3 + \lambda_5 < 8 \pi, ~~~ \lambda_3- \lambda_5 < 8 \pi, \\
    \lambda_1 + \lambda_2 \pm \sqrt{(\lambda_1 -\lambda_2)^2 + \lambda^2_4} < 8 \pi, ~~~  \lambda_1 + \lambda_2 \pm \sqrt{(\lambda_1 -\lambda_2)^2 + \lambda^2_5} < 8 \pi, \\
    \lambda_3 + 2 \lambda_4 \pm 3 \lambda_5 < 8 \pi, ~~2 \lambda_{S} < 8 \pi, ~~\lambda_{S1} < 8 \pi, ~~~\lambda_{S2} < 8 \pi .
\end{eqnarray} 
The aforementioned conditions corresponding to charge-neutral scatterings can not be written in a closed analytical form. Hence, we estimate them numerically and provide the required expressions in the Appendix \ref{unicon}. In the limit  $\lambda_s \to 0$, $\lambda_{S1} \to 0$, and $\lambda_{S2} \to 0 $, the above conditions simplify to the most stringent condition of the standard 2HDM~\cite{Branco:2011iw} as shown:
\begin{equation}
    3 (\lambda_1+\lambda_2) \pm \sqrt{9 (\lambda_1-\lambda_2)^2+(2 \lambda_3 +\lambda_4)^2}< 8 \pi
\end{equation}
\subsection{Experimental Constraints}
The 2HDM faces stringent constraints from experimental data from flavor physics, Higgs boson searches, precision measurements of oblique parameters, as well as dark matter and collider constraints from experiments. We discuss each of these constraints below:
\subsection*{Constraints from flavor physics}
Data from the heavy flavor physics experiments are very useful in constraining the 2HDM parameter space. To constrain the model parameter space, we use the
following heavy flavour physics observable: rare B-meson decays ($B \to X_s\,\gamma$~\cite{Lees:2012ym}, $B_s \rightarrow \mu^{+}\mu^{-}$~\cite{Aaij:2013aka,Chatrchyan:2013bka}, $B^{\pm} \rightarrow \tau^{\pm} \nu$~\cite{Belle-II:2025ruy}) and neutral meson mixing depending on the different types of 2HDM.
In Type I 2HDM, low $\tan \beta < 3$ regions are stringently constrained from $B_s \rightarrow \mu^+ \mu^-$ and $\Delta M_s$~\cite{Mahmoudi:2009zx,Enomoto:2015wbn,Arbey:2017gmh}  while for large values of $\tan \beta$ the bounds on charged Higgs masses from flavor physics are relaxed. For Type II 2HDM there exists a stringent bound on the charged Higgs mass of $m_{H^{\pm}} \geq \textcolor{blue}{600}$\,GeV from $Br(B \to X_s \gamma)$~\cite{Arbey:2017gmh,BaBar:2012fqh}, while the bound is weaker for Type I 2HDM and relaxes to $m_{H^{\pm}} \geq  80$ GeV. As the charged Higgs sector in our case remains the same as 2HDM, with no new decay modes, we ensure that the bounds from Type I 2HDM are used for the study.

\subsection*{Constraints from Higgs Boson searches}
The Higgs sector of 2HDM has been extensively searched at colliders, including LEP and LHC. From searches of pair-produced charged Higgs bosons at LEP in the context of 2HDM, masses below $80$ GeV for the Type II scenario and $72.5$ GeV for the Type I scenario are excluded~\cite{ALEPH:2013htx}. However, in Type I 2HDM, $m_{H^{\pm}}<200$ GeV for $\tan \beta<10$ is constrained by $H^{\pm}\rightarrow \tau^{\pm} \nu$ searches~\cite{Arbey:2017gmh}. Therefore, $\tan\beta \geq 10$ is a safe option to evade these constraints of the collider. 

For the neutral Higgses, we ensure that the second CP-even Higgs, $h_2$ is the SM-like Higgs such that $m_{h_2} = 125.25\pm 0.17$ GeV~\cite{ATLAS:2020coj}. The invisible branching of the SM-like Higgs to a dark matter pair is bounded from  ATLAS and CMS as below
 \begin{align*}
  \begin{split}
  {\rm BR}(h_{2} \rightarrow  Z^{\prime}Z^{\prime}) &\leq 0.07^{+0.030}_{-0.022}    \text{  (ATLAS)}\text{~\cite{ATLAS:2023tkt}}
  \\&\leq 0.15 \text{ (CMS)} \text{~\cite{CMS:2023sdw}}.
  \end{split}
  \end{align*} 
For a vector DM as in our case, the invisible decay width of the SM-like Higgs $h_2$ is given by
\begin{eqnarray}
    \Gamma (h_2 \rightarrow Z^{\prime}Z^{\prime}) = \frac{  g^2_x m_{h_2}(\cos \alpha_2 \sin \alpha_3)^2}{  32\pi x^2_{Z^{\prime}}}(1-4x^2_{Z^{\prime}}+12x^4_{Z^{\prime}}) \sqrt{1-4x^2_{Z^{\prime}}} \,\,\, ,
\end{eqnarray}
where $x^2_{Z^{\prime}}=\frac{m^2_{Z^{\prime}}} {m^2_{h_2}}$
We choose appropriate values for the mixing angles, $\sin\alpha_{1,2,3}$, to ensure that the SM-like 125 GeV Higgs invisible branching ratio is below the current limits of the LHC.

We now discuss the constraints from collider searches of the neutral Higgses.  The constraints from direct searches of non-standard Higgses, particularly, the $H\rightarrow \tau \tau$ searches, are the most stringent constraint in Type II 2HDM for large $\tan \beta$~\cite{ATLAS:2024xfb}. However, Type I 2HDM is free from such stringent constraints. Recent results from Run 2 of LHC show constraints for heavy Higgs masses beyond $\sim$250 GeV~\cite{ATLAS:2024xfb} mainly via the heavy Higgs decaying to two SM-like Higgses or via searches of the pseudoscalar Higgs $A$ decaying to a Z boson and SM Higgs. However, lighter non-standard Higgses remain free from these constraints.
Furthermore, throughout the work, we choose the heavier CP-even Higgs $h_3$ to be dominantly singlet-like via the choice of mixing angles, and hence it remains free from mixing angles. 
di-Higgs and $A\rightarrow Zh$ searches.

\color{black}
\subsection*{Constraints from oblique parameters}
The mass differences of the components of the same multiplet can contribute to oblique parameters, leading to deviations from the standard measured values of $S,\,T$ and $U$ where $S = 0.02 \pm 0.1$, $T =0.07\pm0.12$, and $U = 0.00\pm0.09$~\cite{ParticleDataGroup:2020ssz} and the model predictions of the parameters of $STU$ are obtained from~\cite{Grimus:2007if,Grimus:2008nb}. As dark matter analysis is not sensitive to these mass differences, we choose the particles from the same multiplet to be almost degenerate. Therefore, we choose degenerate masses for the non-standard neutral scalar $h_1$, pseudoscalar $A$, and charged Higgs $H^{\pm}$  for the rest of the study.    

In addition to the above, we consider the constraints from dark matter observables. We discuss these constraints and their impact on the model parameter
space in Sec~\ref{sec:DM}. 
\section{Impact of theoretical constraints}\label{sec:imconstraints}
\begin{figure}[t!]
    \centering
    \includegraphics[width=0.6\linewidth]{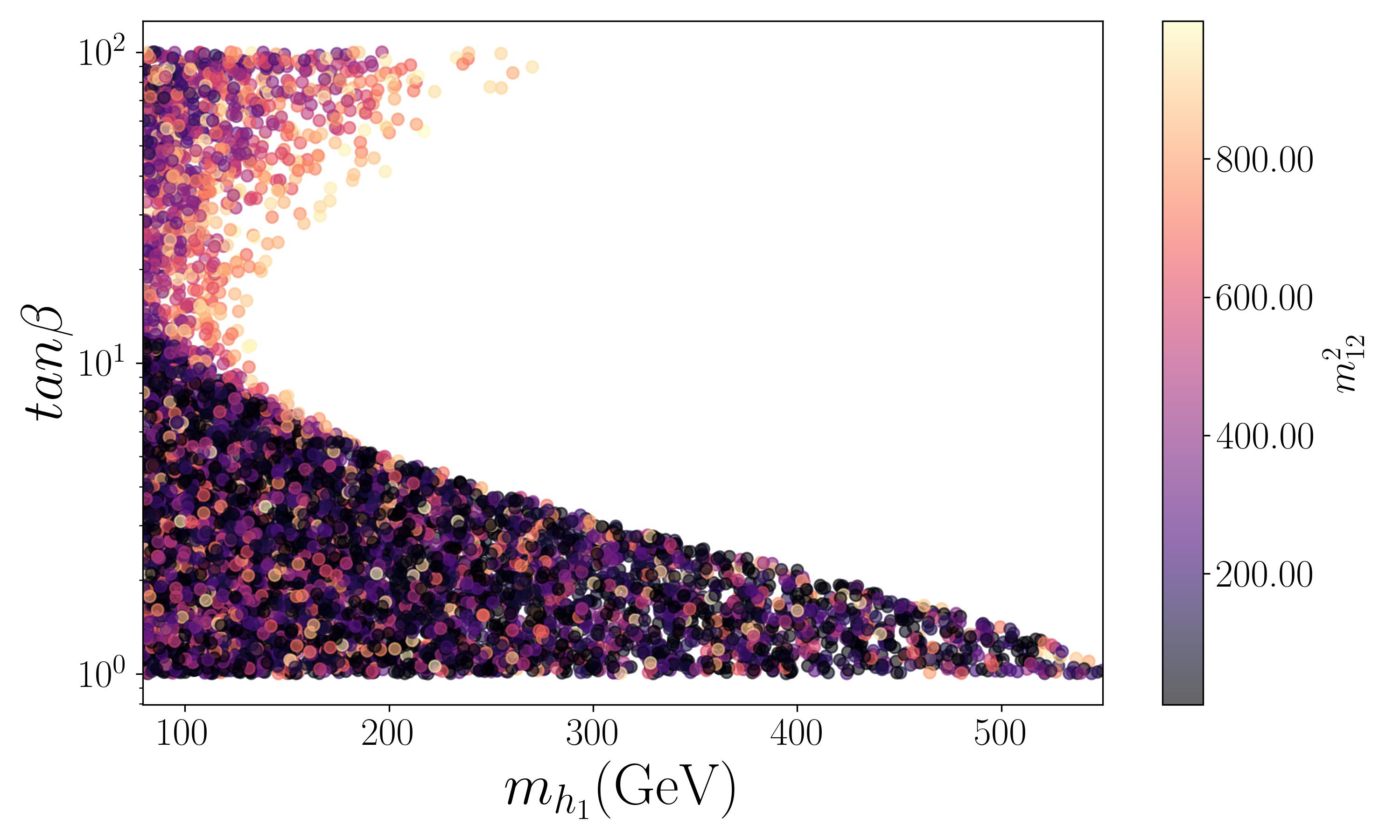}
    \caption{Parameter space in $m_{h_1}$ vs $\tan\beta$ plane allowed from unitarity and BFB conditions. }
    
    \label{fig:unitarity}
\end{figure}

In this section,  we show the effect of theoretical constraints on the model parameter space. Here, we scan over a wide range of Higgs masses without imposing any experimental constraints. Our scan range is as follows 
 \begin{eqnarray*}
    {\rm m_{h_1}}= {\rm m_{h^\pm}} = {\rm m_{A}} &\in& \{80, 1000\},~~~ {\rm m_{h_3}} \in \{ 200, 1000\}, ~~~ \tan\beta \in \{ 1, 100 \}, \nonumber \\
    \sin\alpha_1 &\in& \{0.001, 1 \},~ \sin\alpha_2 \in \{0.001, 1 \},~ \sin\alpha_3 \in \{0.001, 1 \},\nonumber \\
    g_x &\in& \{0.1 , 1 \},~~ {\rm m_{DM}} \in \{10, 1000\}, ~~~{\rm m^2_{12}} \in \{10, 1000\}
\end{eqnarray*}
The parameters of $2$HDM, which are important for dark matter dynamics, are the mass scale of the non-standard Higgs doublet and $\tan\beta$. As the first Higgs model works as a mediator between DM and SM particles, this mass scale gives us control over the light DM mass range. However, the effect of $\tan\beta$ is implicit. Being the ratio of the \textit{vevs} of two Higgs doublets, $\tan\beta$ controls the behavior of the two Higgs bosons when 
$\sin\alpha_i$'s are small. As the couplings of the SM Higgs to the fermions depend on the $\tan\beta$ parameter, the choice of $\tan\beta$  is important. We therefore show the parameter space allowed from theoretical constraints in the $\tan\beta$ vs $m_{h_1}$ plane. We find that, though vacuum stability conditions are not very restrictive, conditions from unitarity exclude a significant amount of the current parameter space. Another important parameter of $2$HDM is $m^2_{12}$. In Fig.\ref {fig:unitarity}, we show the allowed parameter space from unitarity in the aforementioned plane while the corresponding $m^2_{12}$ is shown by the color bar. The maximum value of $m_{h_1}$ is $\sim 700$ GeV and values of $\tan\beta > 10$ are constrained for the mass $m_{h_1}> 100$ GeV. As our dark matter analysis mainly focuses on the lower ranges of dark matter, it necessitates a lighter Higgs. Therefore, we choose a mass range of $80 - 120$ GeV for $m_{h_1}$ and $\tan\beta=10$.

\section{Dark Matter Phenomenology} \label{sec:DM}
\begin{figure}[t!]
    \centering
    \includegraphics[width=0.8\linewidth]{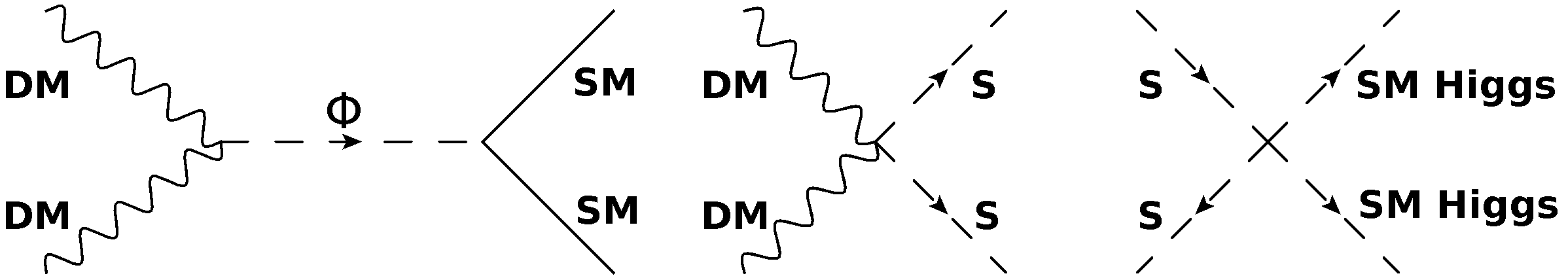}
    \caption{ Diagrams for thermalisation of $Z^\prime$. In the leftmost diagram, $\Phi$ denotes the $3$ Higgs bosons $h_1, h_2, h_3 $ in mass eigenstate.  }
    \label{thermalisation}
\end{figure}

Now we move to the dark matter phenomenology of our model. Due to the imposition of an exact dark charge conjugation symmetry, $Z^\prime$ cannot interact with the photon and the $Z$ boson. The only connection it has with SM is through the Higgs portal. However, for DM freeze-out, the necessary condition is that DM has to be in thermal equilibrium with the bath in the early universe. The main parameters controlling the thermalization of DM are the cross quartic couplings $\lambda_{S1}, \lambda_{S2}$ and the $U(1)_X$ gauge coupling, $g_x$. Their importance can be understood from the representative diagrams shown in Fig.\ref{thermalisation}. The first two diagrams are mainly based on the gauge coupling, while the last is dictated only by $\lambda_{S2}$. The first diagram corresponds to a process in which the DM-SM particle interaction is mediated by the scalar mass eigenstates ($h_1, h_2, h_3$) and vice versa. The other two diagrams show a process in which DM thermalizes with the $S$ singlet scalar (dominantly $h_3$) for an order one $U(1)_X$ gauge coupling, which interacts further with the SM Higgs by quartic couplings. Our choice of $g_x \in [0-1]$ keeps $Z^\prime$ in equilibrium with $S$. 
To ensure adequate mixing angles for our chosen parameters, we take $\lambda_{S1}$ and $\lambda_{S2}$ to be greater than $10^{-4}$. This choice provides sufficient interaction strength with the Higgs boson to maintain dark matter in thermal equilibrium.
Ensuring the thermalisation of DM in the current scenario, we explore the parameter space of our model and its detection aspects that satisfy the relic density constraint. 
\subsection{Relic constraint and Direct Detection}
The only thing we know about dark matter so far, based on the $\Lambda$CDM model, is its current density. The comoving number density of our considered DM species is governed by the following Boltzmann equation 
\begin{equation*}
    \frac{dY_{Z^\prime}}{dx}= \langle \sigma v \rangle _{Z^\prime Z^\prime \to X Y} \frac{\beta s}{\mathcal{H} x} (Y^2_{Z^\prime}-{Y_{Z^\prime}^{eq}}^2 )
\end{equation*}.
Here $Y_{Z^\prime}$ is defined as $n_{Z^\prime}/s$, the ratio of the number density of $Z^\prime$ to the entropy density ($s$) of the visible sector. $\langle \sigma v \rangle _{Z^\prime Z^\prime \to X Y}$ is the velocity averaged cross section of the process $ Z^\prime Z^\prime \to X Y$ where $X, Y$ stand for SM particles and the additional non-standard Higgses. $x=m_{h}/T$ where $T$ is the temperature and $\beta = (1+ \frac{1}{3} \frac{d ln g_s}{d ln T})$ where $g_s$ is the degree of freedom associated with entropy. The expression of the Hubble expansion rate is given by
\begin{equation}
\mathcal{H}= \sqrt{\frac{\pi^2 g_{* \rho}}{90}} \frac{T^2}{M_{Pl}}
\end{equation}
where $g_{* \rho}$ is the total number of relativistic degrees of freedom at temperature $T$ and $M_{Pl}=2.4 \times 10^{18}$ GeV. The relic density ($\Omega h^2$) and the direct detection cross-section for $Z^\prime$ have been calculated using {\tt micrOMEGAs}. We have implemented the model in {\tt FeynRules} and used the generated {\tt calcHEP}~\cite{Belyaev:2012qa}  files in {\tt micrOMEGAs (v5.3.41)}~\cite{Belanger:2006is}.

\begin{figure}[t!]
    \centering
    $$
    \includegraphics[width=0.5\linewidth]{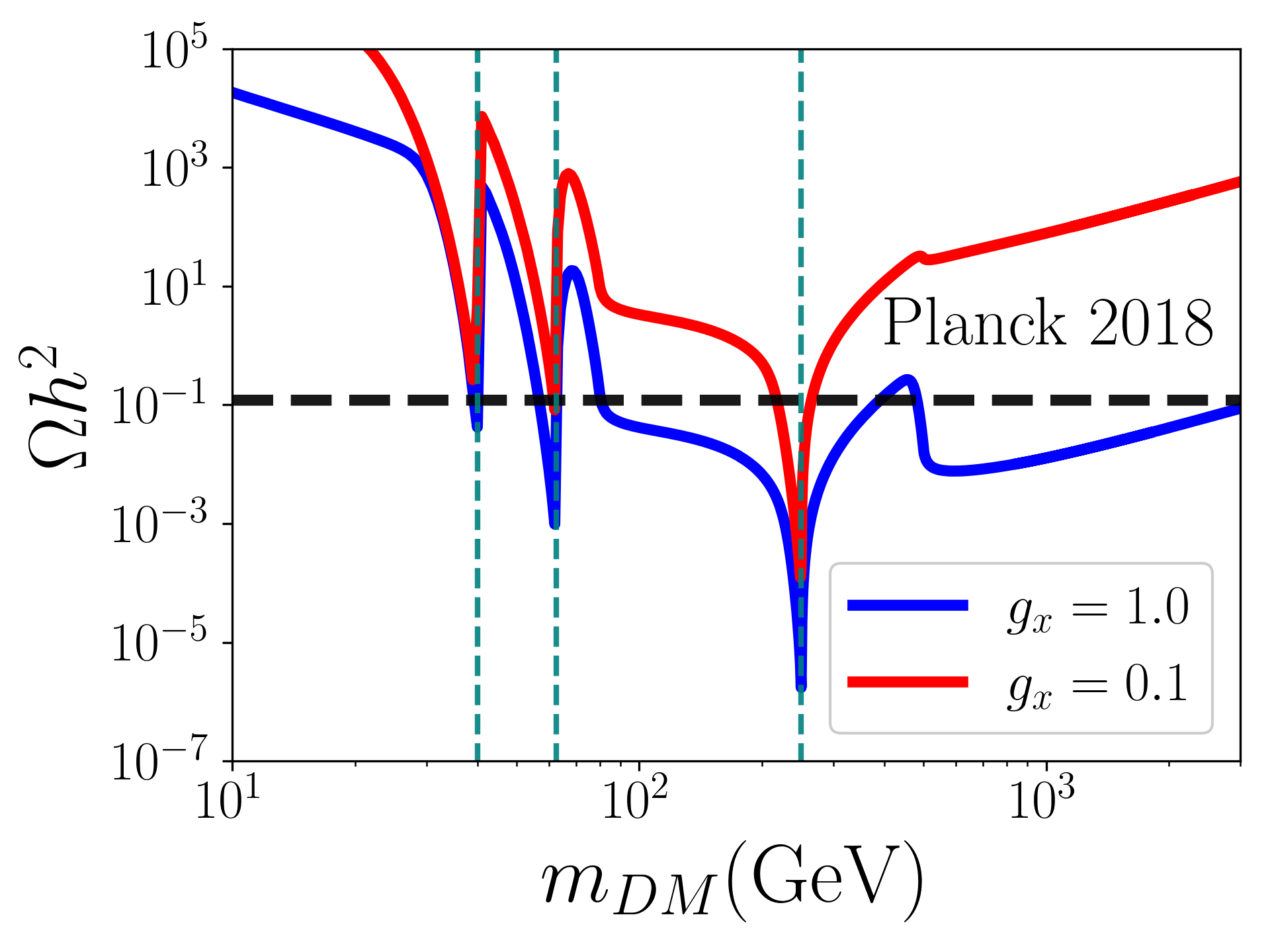}~
    \includegraphics[width=0.5\linewidth]{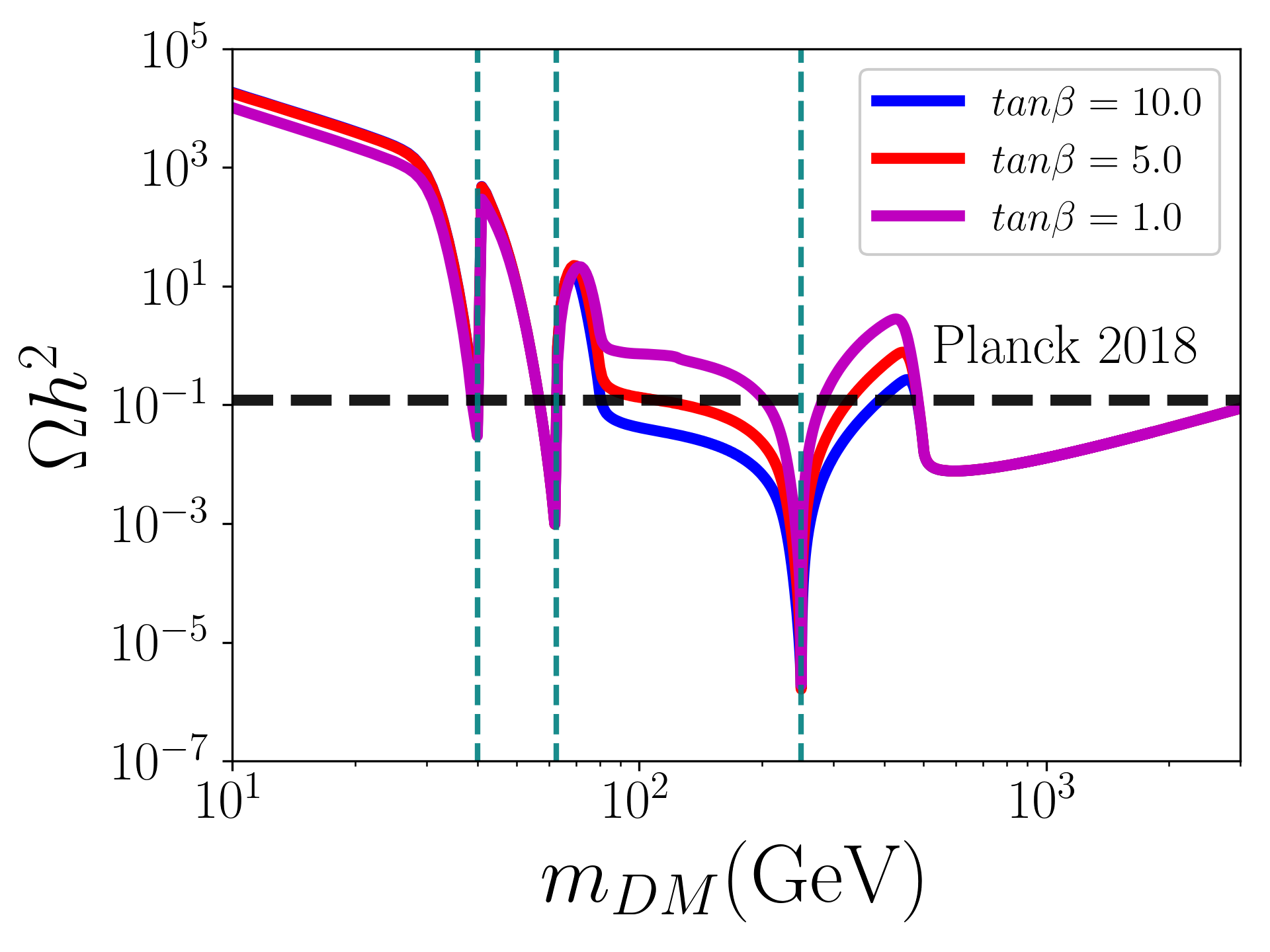}
    $$
    \caption{Relic abundance vs ${\rm m}_{DM}$ (GeV). The Higgs masses are fixed at $m_{h_1}=80 \,{\rm GeV},m_{h_2}=125 \,{\rm GeV}, m_{h_3}=500\,{\rm GeV}$. In the left panel, two values of $g_x$ are considered when $tan\beta=10$. In the right panel, three values of $\tan\beta$ are considered when $g_x=1.0$. The other parameters are fixed at $\sin\alpha_{1}=0.01$, $\sin\alpha_{2}=0.01$ and $\sin\alpha_{3}=0.01$. }
    \label{fig:lineplt}
\end{figure}

In the left panel of Fig. \ref {fig:lineplt}, we show the variation in relic density as a function of DM. For this, we fix three Higgs masses $(m_{h_1},m_{h_2}, m_{h3})$ at $80$ GeV, $125$ GeV, and $500$ GeV, respectively. In addition to the SM Higgs mass, the two other masses are two representative values of the lighter and heavier mass regions compared to the SM Higgs mass. Since the annihilation channels are mediated by the Higgs boson, the corresponding resonances appear near $\sim m_{h}/2$, leading to a sharp drop in the relic density. The black dashed line shows the relic density measured by {\tt PLANCK} \cite{Planck:2018vyg}. The red and blue lines show the relic abundance for $g_{x}=0.1$ and $g_{x}=1.0$, respectively. Increased coupling results in a higher annihilation rate; therefore, more underabundant parameter space is presented, which is shown in the figure. Except for resonances, the drop in relic density after $m_{DM}=80$ GeV and $m_{DM}=500$ GeV is due to the opening of channels $Z^\prime Z^\prime \to h_{1} h_{1}$ and $Z^\prime Z^\prime \to h_{2} h_{2}$, respectively.   In the right panel of Fig. \ref {fig:lineplt}, the $\tan\beta$ dependence of relic density has been shown. With increasing $\tan\beta$, the decay width of $h_3 (S)$ increases. This happens because the relevant Higgs trilinear couplings increase with $\tan\beta$, thereby increasing DM annihilation cross-section ($ Z^\prime Z^\prime \to h_i h_j$). As a result, the under-abundant region increases with increasing $\tan\beta$. Moreover, lower values of $\tan\beta$ alter the decay width of the SM Higgs. Therefore, we fix $\tan\beta=10$ to strengthen the phenomenology of DM and reduce the impact of Higgs constraints.    

\begin{figure}[t!]
    \centering
    \includegraphics[width=0.45\linewidth]{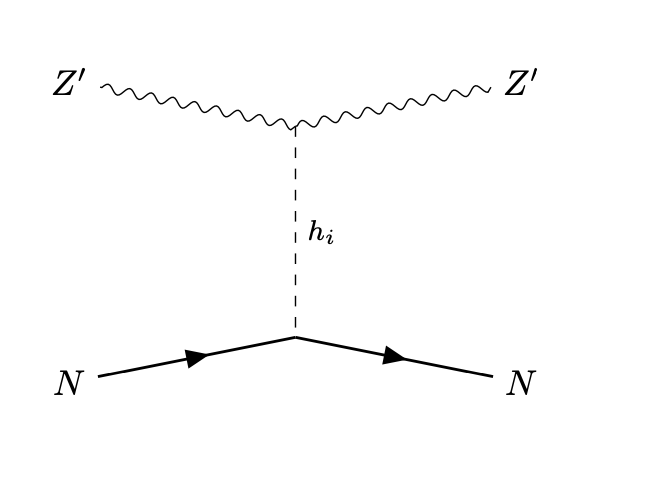}
    \caption{The Feynman diagram corresponding to the direct detection of the dark matter candidate $Z^{\prime}$. Here, $h_i$ ($i=1,2,3$) refer to the Higgs bosons and $N$ refers to a nucleon (proton or neutron).}
    \label{fig:ddfd}
\end{figure}

\begin{figure}[t!]
    \centering
    \includegraphics[width=1.0\linewidth]{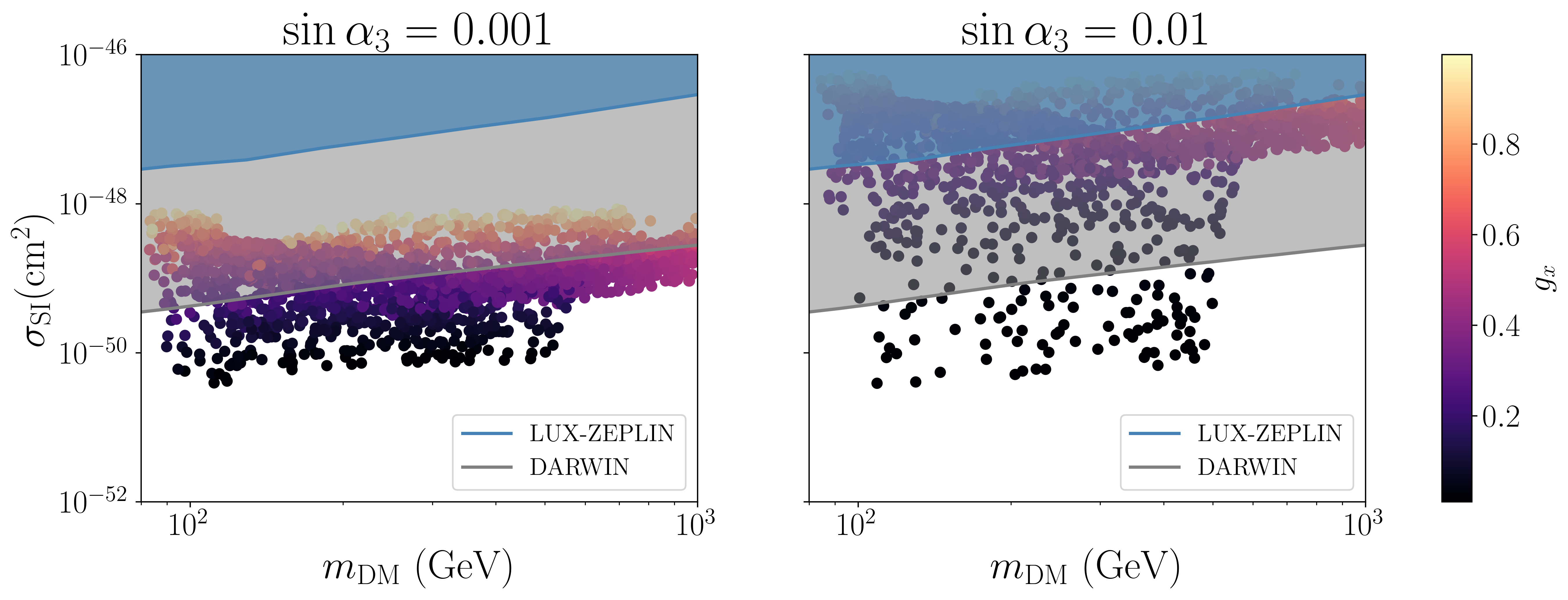}
    \caption{Parameter space with correct relic density in $m_{DM}$ vs $\sigma_{SI}$ plane for different values of mixing angles. The other Higgs doublet mass is varied between $m_{h_1} \in (80.0, 110.0)$GeV, the singlet Higgs mass is varied between $m_{h_3} \in (200, 1000)$ GeV. The other parameters are fixed at the following values:$\tan\beta=10$. For left panel, $\sin\alpha_{3}=0.001$ where as in the right panel, $\sin\alpha_{3}=0.01$.}
    \label{fig:dd1}
\end{figure}

\begin{figure}[b!]
    \centering
    \includegraphics[width=0.6\linewidth]{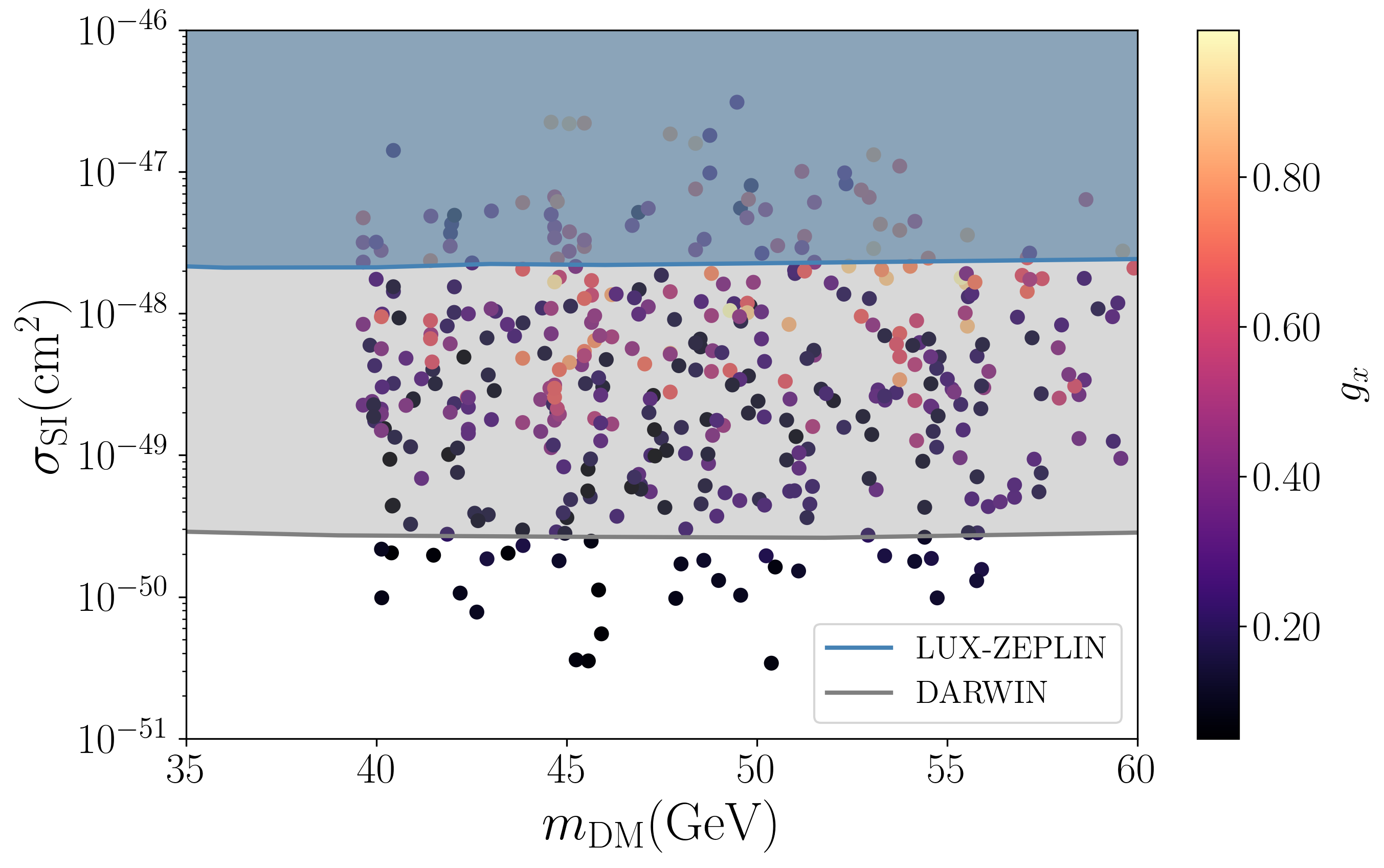}
    \caption{Relic density satisfied parameter space in $m_{DM}$ vs $\sigma_{SI}$ plane for lighter DM mass range.}
    \label{fig:dd}
\end{figure}

After discussing the basic nature of parameters, we now explore the relic density satisfied parameter space in light of DD experiments. The direct detection channel is mediated by Higgs bosons, as seen in Fig~\ref{fig:ddfd}. Direct detection only considers the interaction of DM with nucleons and, therefore, with quarks. In Type I 2HDM, only $\Phi_2$ couples to quarks and $S$ only couples to DM on the gauge basis. As a result, the direct detection cross-section significantly depends on the mixing angle between $\Phi_2$ and $S$, denoted as $\alpha_3$, in addition to the gauge coupling $g_x$. In Fig.\,\ref{fig:dd1}, we show a parameter space in the $m_{DM}$ vs $\sigma_{SI}$ plane that gives the correct relic density within the 2$\sigma$ range, consistent with current DD constraints. The left panel (right panel) corresponds to $\sin\alpha_3=0.001$ ($\sin\alpha_3=0.01$). With a smaller mixing angle, the cross-section falls proportionally, therefore, we can have a parameter space which can be probed in the far future, while a parameter space with a relatively larger mixing angle ($\sin\alpha_3=0.01$) is highly constrained by the current LZ limit. 
The dependence on $g_x$ is shown by the color bar. Here we have plotted the DD cross section for the DM mass range $m_{DM} \in [80,1000]$ GeV. In the higher mass region ($m_{DM} \in [100,1000]$), DM annihilation can result in correct abundance by any combination of $h_3\,(S)$ resonance and opening of the $Z^\prime Z^\prime \to h_i h_j$ channel. The dark colored points (black) with smaller values of the DD cross-section are dominated by resonance annihilation. As a result, dark points are allowed up to DM mass $500$ GeV as shown in the figure, since the largest Higgs mass considered is $ 1000$ GeV. The light colored points with relatively higher DD cross-sections satisfy the relic as more channels become available due to kinematics. However, in the range $[80, 100]$ GeV, due to the absence of a resonance, the process occurs only by opening the channel $Z^\prime Z^\prime \to h_1 h_1/h_2$. 
In the mass region $\leq 80$ GeV, the SM Higgs resonance corresponds to a DM mass of $62.5$ GeV, which is highly constrained by the Higgs invisible branching ratio. However, the presence of an additional Higgs with mass $ \geq 80$ GeV allows for a viable DM in the mass range $[40,60]$ GeV. In Fig.\,\ref{fig:dd}, we show the spin-independent cross section for the relic density satisfied points in this mass range. The blue shaded region shows the parameter space excluded by the LZ  experiment. The gray shaded region shows the parameter space that can be probed by {\tt DARWIN} soon.  

\begin{table}[]
    \centering
    \begin{tabular}{|c|c|c|c|c|c|} \hline
          No.& $g_{x}$ & $m_{DM}$ (GeV) & $m_{h_{1}}$ (GeV) & $m_{h_{3}}$ (GeV)  & $\sigma_{SI}(cm^2$)   \\ \hline
           BP1 & 0.16 & 54.78 & 109.97 & 300.0 & $4.92 \times 10^{-49}$  \\ \hline
            BP2 & 0.20 & 44.35 & 89.74 & 300.0  & $1.08  \times 10^{-48} $  \\ \hline
             BP3 & 0.28 & 90.50 & 87.28 & 221.71 & $2.27  \times 10^{-48} $  \\ \hline
            BP4 & 0.35 & 177.64 & 101.30 & 530.84 & $6.71  \times 10^{-48} $  \\ \hline
            BP5 & 0.48 & 527.0 & 101.90 & 478.71  & $1.20  \times 10^{-47} $  \\ \hline
       \end{tabular} 
       \begin{tabular}{|c|c|c|c|}
          No. & Annihilation channels & No. & Annihilation channels \\ \hline
          BP1 & $Z^\prime Z^\prime \rightarrow b \bar{b}~(80\%)$ & BP2 &  $Z' Z' \rightarrow b \bar{b}~(80\%)$ \\
              & $Z'Z'\rightarrow c ~\bar{c}~(12\%)$ & &   $Z'Z'\rightarrow c ~\bar{c}~(12\%)$\\
              &  $Z'Z'\rightarrow \tau~\bar{\tau}~(8\%)$ & & $Z'Z' \rightarrow \tau~\bar{\tau}~(8\%)$ \\ \hline
          BP3  & $Z'Z'\rightarrow H^{+} H^{-}~(42\%)$ & BP4 & $ Z'Z'\rightarrow H^{+} H^{-}~(46\%)$ \\
              & $ Z'Z'\rightarrow h_1h_1 ~(35\%)$ & & $ Z'Z' \rightarrow h_1~h_1~(29\%) $ \\
              &  $ Z'Z'\rightarrow A~A~(21\%)$ & & $Z'Z'\rightarrow A~A~(23\%)$ \\ \hline
            BP5 & $ Z'Z'\rightarrow h_{2} h_{2}~(97\%)$ & &  \\
              &  $Z'Z'\rightarrow H^{+} H^{-}~(2\%)$ & &  \\ \hline   
       \end{tabular}
    \caption{Parameter points satisfying correct relic abundance, direct detection, and indirect detection constraints. The values of all the mixing angles ($\sin\alpha_{1},\sin\alpha_{2},\sin\alpha_{3}$) chosen for the benchmark points are $\simeq 0.01$. The relative contribution of the main DM annihilation channels for each benchmark point is also shown.}
    \label{tab:benchmarks}
    
\end{table}

\subsection{Indirect Detection}
\begin{figure}[t!]
    \centering
    \includegraphics[scale=0.5]{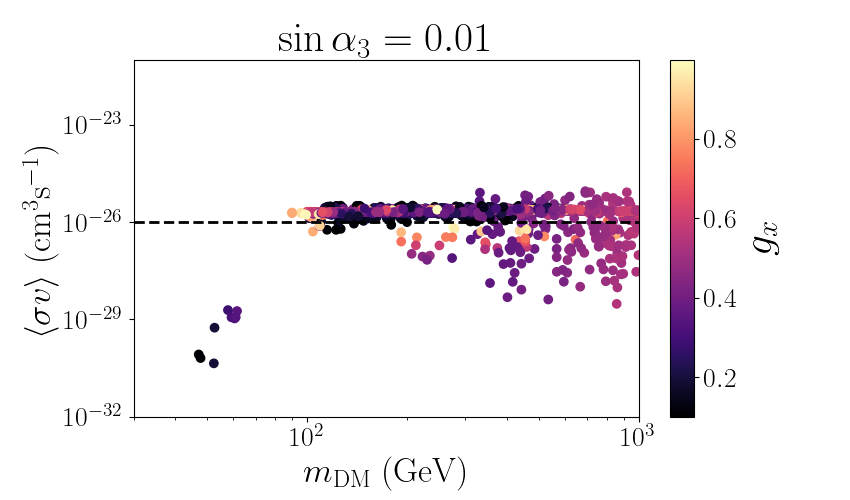}
    \caption{Relic density satisfied parameter space in annihilation cross section vs $m_{DM}$ plane.}
    \label{fig:ind1}
\end{figure}
Dark Matter can also be detected indirectly via its annihilation products in galaxies. The DM pair annihilates in the final SM and BSM states through Higgs-mediated s-channel processes and quartic interactions of DM with Higgses, as shown in Fig.~\ref{thermalisation}. In Fig.~\ref{fig:ind1} we show the variation of the velocity-averaged DM annihilation cross section, $\langle\sigma v\rangle$, versus the dark matter mass $m_{DM}$ with $g_x$ in the colored palette. The plot is done for $\sin \alpha_3=0.01$ with an upper bound of 10$^{-25} cm^3 s^{-1}$ to evade current limits of {\tt Fermi-LAT}. The black dashed line denotes the thermal DM annihilation cross-section 10$^{-26}$ cm$^{-3}s^{-1}$ which corresponds to the required annihilation cross-section for the correct relic density.

We observe that for $m_{DM}<100$ GeV, the total cross-section of DM annihilation $\langle\sigma v\rangle$ is always less than $10^{-26} cm^3s^{-1}$.  
For low DM masses, the dominant annihilation channels near the Higgs resonance regions are the SM channels, i.e, $b\bar{b},$ $c\bar{c},\tau^+\tau^-$. For $m_{DM}>90$ GeV, the bosonic channels $W^+W^-, \, ZZ$ open along with $H^+H^-, \, h_2 \,h_2$ and $A \,A$. Although the SM branching modes are large near the Higgs resonance, particularly for $b\bar{b}$, it is safe from {\tt Fermi-LAT} limits due to the low value $\langle\sigma v\rangle$. For heavier DM masses, the BSM channels $AA, H^+H^-, W^+H^-$ $(i,j=1,2)$ dominate over the SM channels and are safe from indirect detection limits from {\tt Fermi-LAT}, since these channels are not constrained directly from the data.   

To show the effect of the light Higgs doublet in the dark matter parameter space, we list five benchmark points that are consistent with correct dark matter relic abundance ($2\sigma$ limit), current DM direct detection and indirect detection limits 
 in Table.\ref{tab:benchmarks}. For the first two benchmark points,
 where the dark matter mass lies in the range of [40-60] GeV, the resonant annihilation through the light Higgs doublet is possible. This is an important feature of our scenario. The next two benchmark points are representative of the parameter space where $m_{DM} < m_{h_2}$ and the dark matter always dominantly annihilates to the light scalars. In the above-mentioned parameter spaces, the light Higgs doublet plays a significant role in the dark matter phenomenology. In the region where $m_{DM} > m_{h_2}$, the dark matter annihilates to a pair of $H_2$s mimicking the phenomenology of the conventional minimal VDM scenario. In the second part of the table, the annihilation channels of the dark matter with relative percentage have been shown. 
\section{Discussion on Freeze-in prospect}
\label{sec:Freezein}
\begin{figure}[t!]
    \centering
    \includegraphics[width=0.5\linewidth]{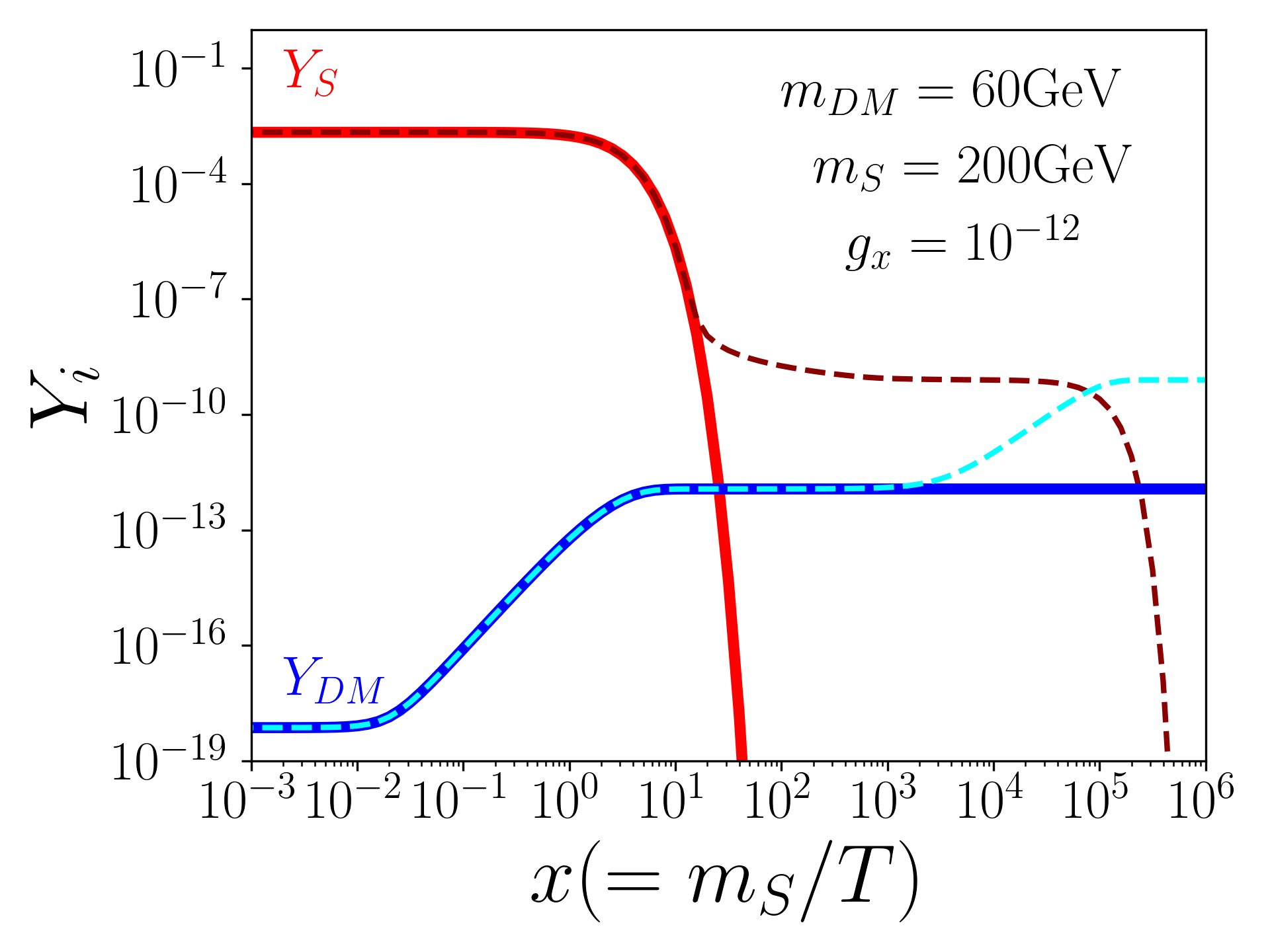}
    \caption{Evolution of comoving number density of DM and parent particle $S$. The dashed (solid) line corresponds to the scenario where decay of $S \to \Phi_1 \Phi_1$ is absent (present).}
    \label{DM evolution}
\end{figure}

While an ${\cal O}(1)$ gauge coupling makes the freeze-out production mechanism a natural candidate for DM generation, the continued non-observation of DM till recent times motivates us to look for a complementary approach. Assuming a very tiny gauge coupling for our case, one can still achieve a correct density of DM non-thermally \cite{Hall:2009bx}. Therefore, even with a negligible initial abundance, $Z^\prime$ can still be produced through the decay and scattering of $S$ ($h_3$) particles. As we shall consider very small mixing angles, the gauge and mass eigenstates will be almost the same ($S\sim h_3, \Phi_1 \sim h_1, \Phi_2 \sim h_2$), and therefore we will mention them interchangeably. $S$ can be initially in thermal equilibrium with the SM bath via the quartic couplings with two Higgs doublets. Being directly coupled to $S$, the couplings of $Z^\prime$ to $\Phi_1$  and $\Phi_2$ are $\sin\alpha_i$ suppressed; therefore, their contribution to $Z^\prime$ abundance can be neglected. The relevant number density for the freeze-in scenario would be that of the $S$ particle and DM $Z^\prime$. We solve the two coupled Boltzmann equations mentioned in the appendix \ref{BEQ} to study the evolution of DM density. In Fig.\,\ref{DM evolution}, we present the evolution of comoving number densities of the above two species. The dashed lines correspond to a scenario in which the extra light  Higgs doublet is not present in the spectrum. Therefore, $S$ decays only in DM as the other decay channels in SM particles are suppressed by the mixing angle. Initially, $S$ remains in thermal equilibrium through its interaction with the SM Higgs. After some time, it freezes out of equilibrium but finally transfers its number density to the DM density, as shown in the figure. Here, the mass of $S$ is such that its decay to two SM Higgs is kinematically forbidden. However, having $\Phi_1$ as the light Higgs doublet in the spectrum provides an extra decay channel for $S$, thus keeping $S$ in thermal equilibrium throughout the period. 
This feature is visible in the solid lines of Fig.\,\ref{DM evolution}. In the parameter space available, where $m_{S}<2 m_{\Phi_2}$, the presence of a light $\Phi_1$ can stop freezing out of $S$, therefore keeping the relic density in the right range. This can be considered as an advantage of having an extra light Higgs doublet in the spectrum. Therefore, we have safely considered $S$ to be in thermal equilibrium, following which the Boltzmann equation of the DM number density can be written as 
\begin{equation}
    \frac{dY_{\rm DM}}{dx} = \frac{\langle \Gamma _{S \to \rm DM~ DM }\rangle}{\mathcal{H} x} Y^{eq}_S(x)
\end{equation}
where $x=m_S/T$. The DM number density in the present era is given by
\begin{equation}
    \Omega_{\rm DM} h^2 \simeq 2.755 \times 10^8 \, \,
\frac{m_{\rm DM}}{\rm GeV} \, Y^{\rm today}_{\rm DM}  \,\, .
\end{equation}

\begin{figure}[t!]
    \centering
    $$
    \includegraphics[width=0.5\linewidth]{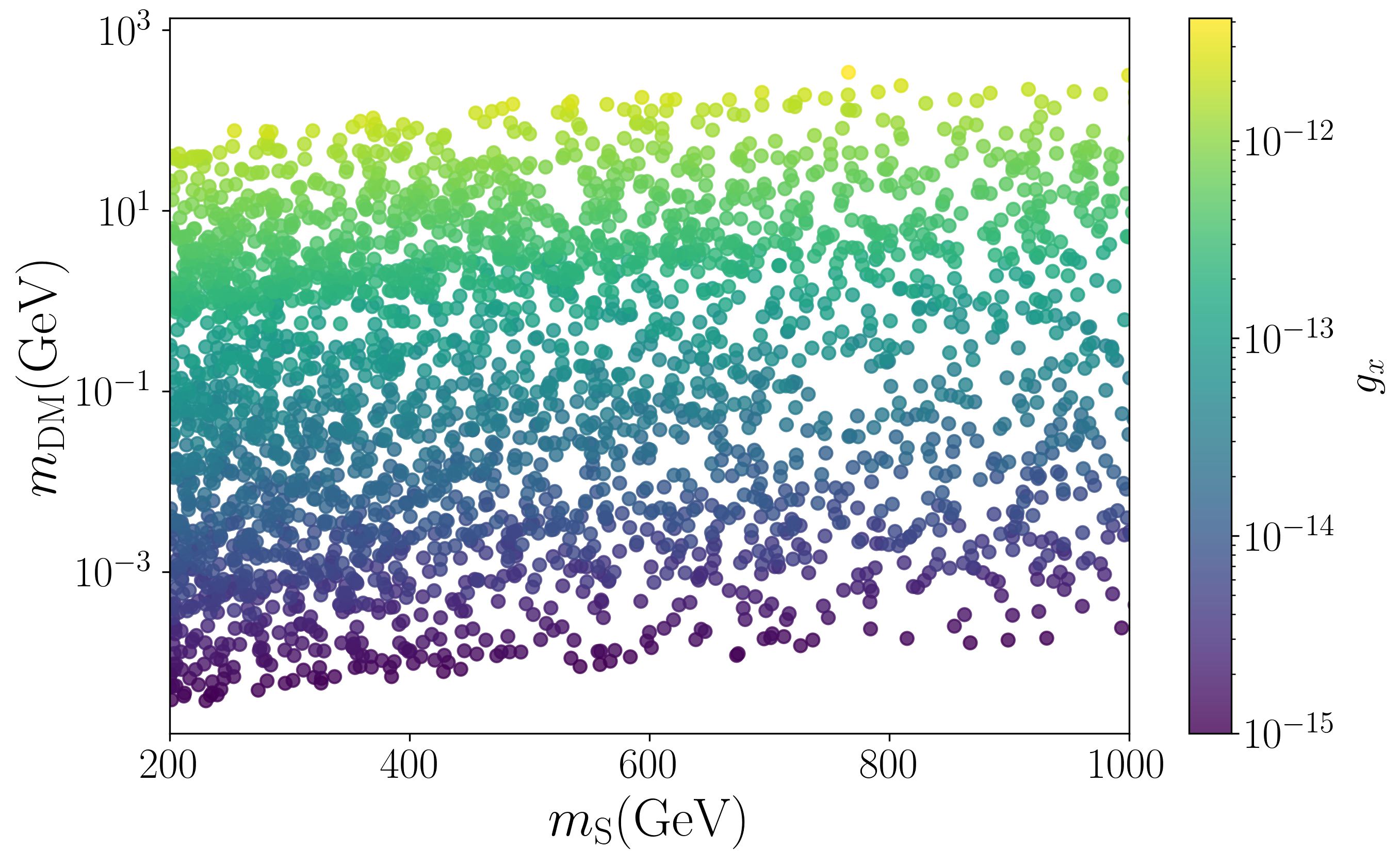}~~
    \includegraphics[width=0.5\linewidth]{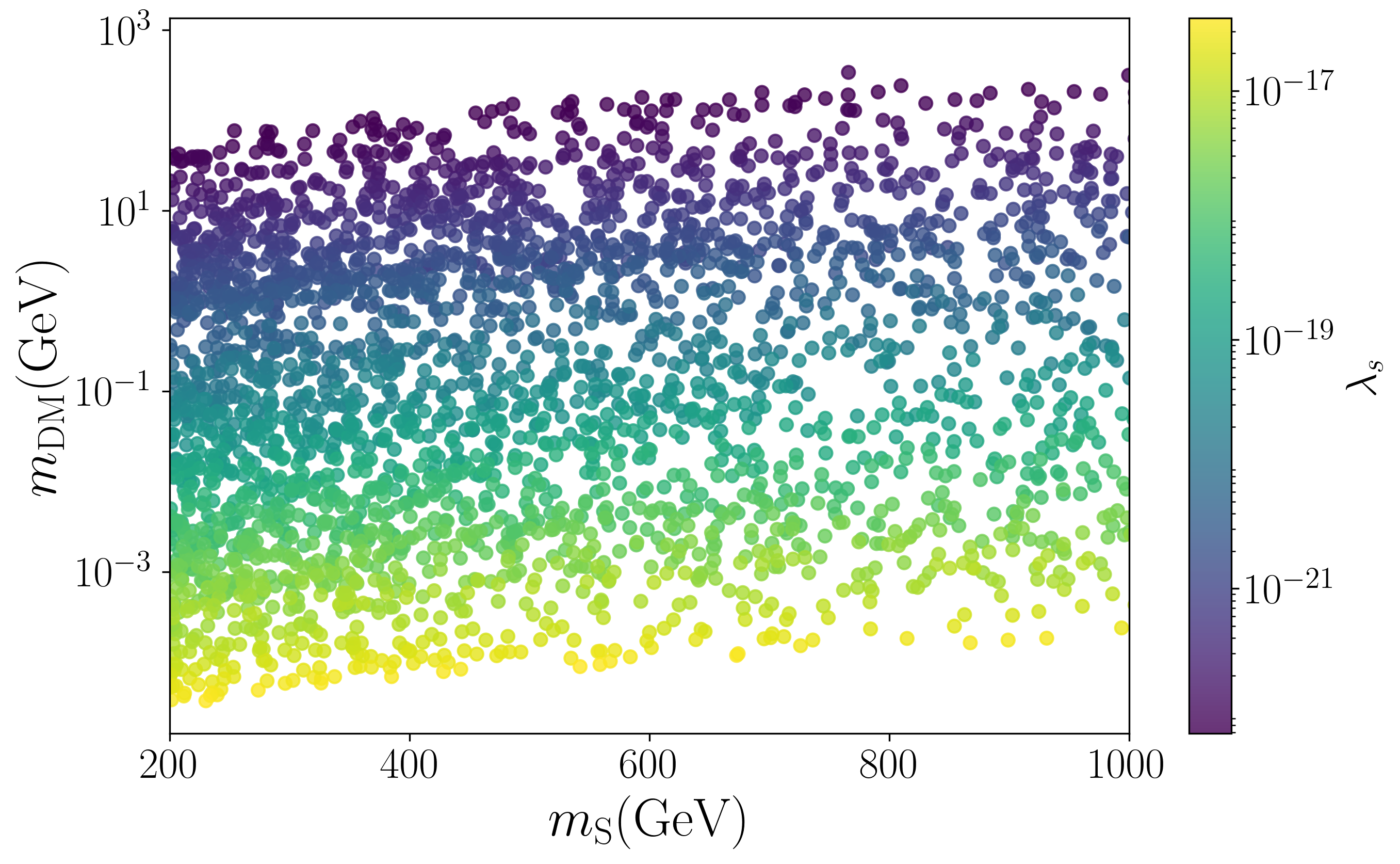}
    $$
    \caption{Parameter space with correct relic abundance in $m_{\rm DM}$ vs $m_{S}$ plane. In the left (right) panel, the variation of gauge coupling $g_x$ (self-coupling of $S$, $\lambda_S$) has been shown. }
    \label{fig:DMscanfreezein}
\end{figure}

The approximate expression of the DM mass and the mother particle S can be expressed as
\begin{equation}
    m_{DM}= g_x v_S, ~~~~~ m_S \sim \sqrt{\lambda_S} v_S \,\, .
\end{equation}
Combining these two equations, we have an approximate relation between the masses of the daughter DM and the mother particle, referred to below
\begin{equation}
    m_{DM} \sim \frac{m_S g_x}{\sqrt{\lambda_S}} \,\, .
\end{equation}
Here, we have performed a random scan to see what values of gauge coupling $g_x$ and $\lambda_S$ can produce a correct relic abundance in the early universe. 
In Fig.\,\ref{fig:DMscanfreezein}, we show the parameter space available with correct relic abundance in the $m_{DM}$ vs $m_S$ plane. The corresponding dependence on $g_x$ and $\lambda_s$ is shown in the color bars in each corresponding plot. As expected, lower 
values of $g_x$ lead to a lighter DM mass. Here we show the DM mass to be up to a few hundred keV. However, we can reduce it to a minimum of  $\sim 5$ keV of DM mass. A lower mass than that would be constrained from free-streaming length simulations \cite{Gilman:2019nap, PhysRevD.109.043511}. To satisfy the relic constraint, DM with mass $5$ keV requires a $g_x$ value of $5.495 \times 10^{-16}$ ($2.512 \times 10^{-16}$) for $m_S=200$ GeV ($1000$ GeV). We can consider these values to be the lowest values of $g_x$ for the freeze-in production of dark matter.    
\section{Summary and Conclusions}
\label{sec:summary}
In this work, we have explored the DM implications of the Universe in the context of 2HDM. In a local $U(1)_X$ extended version of 2HDM, the extra gauge boson $Z^\prime$ can play the role of DM, where we have considered an exact dark charge conjugation symmetry to ensure the stability of the dark matter candidate. In addition to the standard two Higgs doublets, there is an extra complex scalar that is charged under local $U(1)_X$. When this SM singlet complex scalar acquires a \textit{vev}, the $U(1)_X$ symmetry is spontaneously broken, and the dark matter (DM) gains mass by eating up the CP odd degree of freedom of the $U(1)_X$ scalar. The DM interacts with the SM particles via the Higgs portal terms. We investigate the possibility of freeze-out of such a DM candidate and check its detectability under different direct detection and indirect detection experiments. In our analysis of the DM phenomenology in our model, we have considered all possible theoretical and experimental constraints applicable to this scenario. Our choice of 2HDM parameters $\tan\beta$ and $m_{h_1}$ (mass of the non-standard Higgs doublet) is motivated by experimentally favored parameter space. In terms of DM analysis, the thermalization of the DM in the early Universe is ensured by choosing an appropriate strength of interactions with the SM spectrum, which in our case gives $\lambda_{S2} \sim 10^{-4}$ and $g_x \in [0,1]$. The DM number density in the current epoch is calculated using {\tt micrOMEGAs}. The parameter space of our model, which satisfies the correct abundance of relic predicted by {\tt PLANCK} within 2$\sigma$,  
can be probed in future DD experiments. A key feature of our model is the existence of a viable parameter space in the dark matter mass range of $[40 - 60]$ GeV, even under strong direct detection limits.  Generally, due to the resonance of the SM Higgs, some models allow dark matter masses around $62.5$ GeV, which is severely restricted by the invisible branching ratio. In contrast to existing DM scenarios in the context of 2HDM, where DM densities are either underabundant (inert Higgs doublet models) in the mass range $\leq 90$ GeV or very heavy (TeV mass), here we found a reasonable range of DM parameter space in the above mass range that is still allowed by DD experiments using the non-standard Higgs doublet as a mediator. Such a parameter space for light DM mediated by a Higgs doublet is available for type I of 2HDM. Type II 2HDM does not allow a light scalar spectrum due to stringent flavour constraints. However, lepton-specific and flipped 2HDM can allow at least one light pseudoscalar or scalar corresponding to the Higgs doublet for a specific configuration, which can eventually result in a light DM parameter space. But the collider constraints of such types would need careful consideration. Further, we also explore the case for tiny values of the  gauge coupling ($g_x$) where the dark vector boson may become a non-thermal DM candidate. We present the region of allowed parameter space accommodating such a very light DM, with masses in the few keV to GeV region, produced via the freeze-in mechanism and satisfying the observed relic abundance.

\section*{Acknowledgements}
ND would like to thank Anindya Datta, Rohan Pramanick, Tanmoy Kumar, and Koustav Mukherjee for discussions. ND acknowledges financial support from the CSIR, Government of India, through the NET Senior Research Fellowship (SRF), under 
Award File No. 09/080(1187)/2021-EMR-I. ND also acknowledges the hospitality of the Harish Chandra Research Institute, Prayagraj, India, where part of this work was carried out. JD acknowledges support from the OUHEP cluster at the University of Oklahoma and the Institute of Mathematical Sciences, India. SKR acknowledges support from the Department of Atomic Energy (DAE), India, for the Regional Center for Accelerator-based Particle Physics (RECAPP), Harish Chandra Research Institute.

\section{Appendix}
\subsection{Quartic couplings in terms of physical parameters}
\label{qcoup}
The quartic couplings can be written in terms of physical parameters (mass, mixing angles, and \textit{vev}) as 
\begin{eqnarray}
    \lambda_1 &=& \frac{1}{2 v^2_1} \{ m^2_{h_1} \cos\alpha^2_1 \cos\alpha^2_2 + m^2_{h_2} (\cos\alpha_3 \sin\alpha_1 +\cos\alpha_1 \sin\alpha_2 \sin\alpha_3)^2+ m^2_{h_3} (\sin\alpha_1 \sin\alpha_3  \nonumber \\
 && -\cos\alpha_1 \cos\alpha_3 \sin\alpha_2)^2\} - m^2_{12} \frac{v_2}{2 v^3_1}   \\
   \lambda_2 &=& \frac{1}{2 v^2_2} \{ m^2_{h_1} \sin\alpha^2_1 \cos\alpha^2_2 + m^2_{h_2} (\cos\alpha_1 \cos\alpha_3 - \sin\alpha_1 \sin\alpha_2 \sin\alpha_3)^2 + m^2_{h_3} (\cos\alpha_1 \sin\alpha_3 \nonumber \\
   &&+ \sin\alpha_1 \sin\alpha_2 \cos\alpha_3)^2 \} -m^2_{12} \frac{v_1}{2 v^3_2} \nonumber \\
   \lambda_3 &=& \frac{1}{v_1 v_2} \{ m^2_{h_1} \sin\alpha_1 \cos\alpha_1 \cos\alpha^2_2 + m^2_{h_2} (\cos\alpha_1 \sin\alpha_1 \sin\alpha^2_2 \sin\alpha^2_3 + \cos\alpha_3 \sin\alpha_2 \sin\alpha_3 \sin\alpha^2_1 \nonumber \\
   && -\sin\alpha_2 \sin\alpha_3 \cos\alpha_3 \cos\alpha^2_1 - \sin\alpha_1 \cos\alpha_1 \cos\alpha^2_3) + m^2_{h_3} (\cos\alpha_1 \cos\alpha^2_3 \sin\alpha_1 \sin\alpha^2_2  \nonumber \\
   && + \cos\alpha^2_1 \cos\alpha_3 \sin\alpha_2 \sin\alpha_3 - \cos\alpha_3 \sin\alpha^2_1 \sin\alpha_2 \sin\alpha_3 -\cos\alpha_1 \sin\alpha_1 \sin\alpha^2_3)-m^2_{12}\}\nonumber \\&& +\frac{2 m^2_{H^{\pm}}}{(v^2_1+ v^2_2)}\nonumber \\
   \lambda_4 &=& \frac{m^2_{12}}{v_1 v_2}+ \frac{m^2_A -2 m^2_{H^{\pm}}}{v^2_1+ v^2_2} \nonumber \\
   \lambda_5 &=& \frac{m^2_{12}}{v_1 v_2} - \frac{m^2_A}{v^2_1+ v^2_1} \nonumber \\
   \lambda_{s} &=& \frac{1}{2 v^2_S} (m^2_{h_1} \sin\alpha^2_2 + m^2_{h_2} \cos\alpha^2_2 \sin\alpha^2_3 + m^2_{h_3} \cos\alpha^2_2 \cos\alpha^2_3) \nonumber \\
   \lambda_{S1} &=& \frac{1}{v_1 v_S} \{ m^2_{h_1} \cos\alpha_1 \cos\alpha_2 \sin\alpha_2 - m^2_{h_2} (\cos\alpha_2 \cos\alpha_3 \sin\alpha_1 \sin\alpha_3 + \cos\alpha_1 \cos\alpha_2 \sin\alpha_2 \sin\alpha^2_3) \nonumber \\
   &&+ m^2_{h_3} (\cos\alpha_2 \cos\alpha_3 \sin\alpha_1 \sin\alpha_3 -\cos\alpha_1 \cos\alpha_2 \cos\alpha^2_3 \sin\alpha_2)\} \nonumber \\
   \lambda_{S2} &=& \frac{1}{v_2 v_S} \{ m^2_{h_1} \cos\alpha_2 \sin\alpha_1 \sin\alpha_2 + m^2_{h_2} (\cos\alpha_1 \cos\alpha_2 \cos\alpha_3 \sin\alpha_3 - \cos\alpha_2 \sin\alpha_1 \sin\alpha_2 \sin\alpha^2_3) \nonumber \\
   &&- m^2_{h_3} (\cos\alpha_1 \cos\alpha_2 \cos\alpha_3 \sin\alpha_3 + \cos\alpha_2 \cos\alpha^2_3 \sin\alpha_1 \sin\alpha_2) \} \nonumber
    \end{eqnarray}

\subsection{Condition for Unitarity}
\label{unicon}
The amplitude matrix of charge neutral scatterings in all possible two particle states are presented below: 
\begin{equation*}
    \mathcal{M}=\begin{pmatrix}
 2 \lambda_{1} & \lambda_{3}+\lambda_{4} & \frac{\lambda_{1}}{2} & \frac{\lambda_{3}}{2} & \frac{\lambda_{1}}{2} &
   \frac{\lambda_{3}}{2} & \frac{\lambda_{s1}}{2} & \frac{\lambda_{
   s1}}{2} \\
 \lambda_{3}+\lambda_{4} & 2 \lambda_{2} & \frac{\lambda
   _{3}}{2} & \frac{\lambda_{2}}{2} & \frac{\lambda_{3}}{2} &
   \frac{\lambda_{2}}{2} & \frac{\lambda_{s2}}{2} & \frac{\lambda
   _{s2}}{2} \\
 \lambda_{1} & \lambda_{3} & \frac{3 \lambda_{1}}{2} & \frac{1}{2}
   (\lambda_{3}+\lambda_{4}+\lambda_{5}) & \frac{\lambda
   _{1}}{2} & \frac{1}{2} (\lambda_{3}+\lambda_{4}-\lambda_{5}) &
   \frac{\lambda_{s1}}{2} & \frac{\lambda_{s1}}{2} \\
 \lambda_{3} & \lambda_{2} & \frac{1}{2} (\lambda
   _{3}+\lambda_{4}+\lambda_{5}) & \frac{3 \lambda_{2}}{2} &
   \frac{1}{2} (\lambda_{3}+\lambda_{4}-\lambda_{5}) &
   \frac{\lambda_{2}}{2} & \frac{\lambda_{s2}}{2} & \frac{\lambda
   _{s2}}{2} \\
 \lambda_{1} & \lambda_{3} & \frac{\lambda_{1}}{2} & \frac{1}{2}
   (\lambda_{3}+\lambda_{4}-\lambda_{5}) & \frac{3 \lambda
   _{1}}{2} & \frac{1}{2} (\lambda_{3}+\lambda_{4}+\lambda_{5}) &
   \frac{\lambda_{s1}}{2} & \frac{\lambda_{s1}}{2} \\
 \lambda_{3} & \lambda_{2} & \frac{1}{2} (\lambda
   _{3}+\lambda_{4}-\lambda_{5}) & \frac{\lambda_{2}}{2} &
   \frac{1}{2} (\lambda_{3}+\lambda_{4}+\lambda_{5}) & \frac{3
   \lambda_{2}}{2} & \frac{\lambda_{s2}}{2} & \frac{\lambda
   _{s2}}{2} \\
 \lambda_{s1} & \lambda_{s2} & \frac{\lambda_{s1}}{2} &
   \frac{\lambda_{s2}}{2} & \frac{\lambda_{s1}}{2} & \frac{\lambda
   _{s2}}{2} & 3 \lambda_{s} & \lambda_{s} \\
 \lambda_{s1} & \lambda_{s2} & \frac{\lambda_{s1}}{2} &
   \frac{\lambda_{s2}}{2} & \frac{\lambda_{s1}}{2} & \frac{\lambda
   _{s2}}{2} & \lambda_{s} & 3 \lambda_{s} \\
\end{pmatrix}
\end{equation*}
By demanding the largest eigen value of this matrix to be $\le 8 \pi$, we satisfy the conditions of unitarity.
\subsection{Boltzmann equations}
\label{BEQ}
The comoving number density of $S$ and DM are dictated by two coupled Boltzmann equations mentioned below
\begin{eqnarray*}
\frac{dY_{DM}}{dz} &=& \frac{1}{z \mathcal{H}}\langle \Gamma_{S \to DM DM} \rangle Y_{S} (z) +\frac{4 \pi^2}{45} M_{Pl} \frac{m_{S}}{1.66 z^2} \langle \sigma v \rangle_{ S S \to DM DM }Y^2_{S}(z)\\
 \frac{dY_{S}}{dz} &=& - \frac{1}{z\, \mathcal{H}} \langle \Gamma_{S \to DM DM} \rangle Y_{S} (z) - \sum_{i} \frac{1}{z\, \mathcal{H}} \langle \Gamma_{S \to \Phi_{i} \Phi_{i}} \rangle (Y_{S} (z)- Y^{eq}_{S} (z)) \\ 
    &&-\frac{4 \pi^2}{45} M_{Pl} \frac{m_{S}}{1.66 z^2} (\langle \sigma v \rangle_{ S S \to DM DM }Y^2_{S}(z)- \langle \sigma v \rangle_{ S S \to H H }(Y^2_{S}(z)-{Y_{S}^{eq}}^2(z)))
\end{eqnarray*}
where $Y_{i}=\frac{n_i}{s}$ and $i=1,2$.
The thermal average of decay width and scattering cross-section are defined as
\begin{eqnarray*}
\langle \Gamma \rangle &=& \Gamma\, \frac{K_1(z)}{K_2(z)} \\
    \langle \sigma v\rangle &=& \frac{z}{8 m^5_S K^2_2(z)} \int_{4 m^2_S}^{\infty} \sigma(s)(s- 4 m^2_S) \sqrt{s} \, K_1( \frac{\sqrt{s}}{T}) ds
\end{eqnarray*}
 The relevant decay widths are mentioned below
\begin{eqnarray*}
    \Gamma_{S \to Z^\prime Z^\prime} &=&  \frac{1}{32\pi}   \frac{  g^2_x m^3_{S}}{  m^2_{Z^{\prime}}}(1-\frac{4 m^2_{Z^{\prime}}}{m^2_{S}}+\frac{12 m^4_{Z^{\prime}}}{m^4_{S}}) \sqrt{1-\frac{4 m^2_{Z^{\prime}}}{m^2_{S}}} \\
    \Gamma_{S \to\Phi_i \Phi_i} &=&  \frac{(\lambda_{Si} v_S)^2}{8 \pi\,m_S}  \sqrt{1-\frac{4 m^2_{\Phi_i}}{m^2_{S}}} 
\end{eqnarray*}
The relevant cross sections are as follows
\begin{eqnarray*}
    \sigma_{SS \to \Phi_i \Phi_i} &=& \frac{\lambda^2_{Si}}{16 \pi s} \frac{\sqrt{s- 4 m^2_{\Phi_i}}}{\sqrt{s- 4 m^2_S}}\\
    \sigma_{S S \to Z^\prime Z^\prime} &=& \frac{g^4_x}{16 \pi s} \frac{\sqrt{s- 4 m^2_{Z^\prime}}}{\sqrt{s- 4 m^2_S}}(3+ \frac{s^2}{4 m^4_{Z^\prime}}- \frac{s}{m^2_{Z^\prime}})
\end{eqnarray*}
  
\bibliographystyle{JHEP}
\bibliography{ref}

\end{document}